\documentclass[sigconf]{acmart}

\usepackage{amsmath}
	\usepackage{cancel}
\usepackage{mathtools}
\usepackage{graphicx} 
\usepackage{url}
\usepackage{enumitem} 

\usepackage{stmaryrd} 

\usepackage{upgreek} 
\usepackage{bbding} 

\usepackage{listings} 
\usepackage{tabularx} 
\usepackage{booktabs} 
\usepackage{multirow}
\usepackage{hhline} 
\usepackage{diagbox}
\usepackage{pgfplotstable} 

\usepackage{xcolor,colortbl}    

\usepackage{tikz} 
\usetikzlibrary{matrix}
\usetikzlibrary{calc}
\usetikzlibrary{math}
\usetikzlibrary{arrows.meta,positioning}
\usetikzlibrary{intersections, pgfplots.fillbetween}

\usepackage{easybmat} 

\usepackage{adjustbox}
\def\eox{\unskip\kern 10pt{\unitlength1pt\linethickness{.4pt}$\diamondsuit${}}} 

\usepackage{rotating}		

\usepackage{xspace} 

\usepackage{subcaption} 
\newcommand{\revision}[1]{{\color{black}{#1}}}

\newcommand{\hide}[1]{}

\usepackage[ruled,noend,linesnumbered]{algorithm2e} 

\usepackage{setspace} 

\DontPrintSemicolon     
\SetNlSty{}{}{}                
\SetAlgoInsideSkip{smallskip}   

\SetAlFnt{\small}			
\SetAlCapFnt{\small}		
\SetAlCapNameFnt{\small}

\usepackage[capitalise,nameinlink]{cleveref} 

\crefname{section}{Sec.}{Secs.}
\crefname{example}{Ex.}{Exes.}

\hypersetup{                                                            
	pdfpagemode=UseNone,               
    pdfstartview=Fit,
    pdfpagelayout=SinglePage,       
    bookmarks=false,
    bookmarksnumbered=true,
    bookmarksopen=false,             
    pdftex=true,
    unicode,
    colorlinks=true,       
    linkcolor=magenta,          
    citecolor=cyan,  
    filecolor=magenta,      
    urlcolor=blue           
}

\usepackage{aliascnt}  	

\newaliascnt{corollary}{theorem}

\aliascntresetthe{corollary}

\newaliascnt{example}{theorem}
\newtheorem{example}[example]{Example}
\aliascntresetthe{example}

\newaliascnt{definition}{theorem}
\newtheorem{definition}[definition]{Definition}
\aliascntresetthe{definition}

\newaliascnt{proposition}{theorem}

\aliascntresetthe{proposition}

\newaliascnt{lemma}{theorem}

\aliascntresetthe{lemma}

\newaliascnt{conjecture}{theorem}

\aliascntresetthe{conjecture}

\newtheorem{questionW}{Question}
\newtheorem{resultW}{Result}

{\end{itshape}
}
\AtEndPreamble{%
    \hypersetup{colorlinks,
      linkcolor=purple,
      citecolor=blue,
      urlcolor=ACMDarkBlue,
      filecolor=ACMDarkBlue}}

\usepackage{tcolorbox}		
\tcbuselibrary{breakable,skins}		

\tcbset{examplestyle/.style={
		enhanced jigsaw,	
		colback=blue!08,	
		colframe=blue!08,	
		arc=2mm,
		boxrule=0pt,		
		left=1mm,
		right=1mm,
		left skip= 0mm,  
		right skip= 0mm, 
		top=1mm,		
		bottom=1mm,		
		breakable,		
		parbox = false,		
		before={\par\pagebreak[0]\vspace{1mm}\parindent=0pt},		
		after={\par\pagebreak[0]\vspace{1mm}\parindent=0pt},				
		bottomrule = 0mm,
		boxsep = 0mm,					
		topsep at break=0pt,			
		bottomsep at break=0pt,			
		pad at break=0mm,
		pad before break=1mm,		
		pad after break=1mm,		
		bottomrule at break=0mm,
		toprule at break=0mm,		
		}}

\usepackage{footnote}

\tcolorboxenvironment{example}{examplestyle}

\makeatletter
\DeclareRobustCommand*\uell{\mathpalette\@uell\relax}
\newcommand*\@uell[2]{
  \setbox0=\hbox{$#1\ell$}
  \setbox1=\hbox{\rotatebox{10}{$#1\ell$}}
  \dimen0=\wd0 \advance\dimen0 by -\wd1 \divide\dimen0 by 2
  \mathord{\lower 0.1ex \hbox{\kern\dimen0\unhbox1\kern\dimen0}}
}
\makeatother

\usepackage{marginnote}

\newcommand{\introparagraph}[1]{\textbf{#1.}} 

\renewcommand{\epsilon}{\varepsilon} 

\renewcommand{\O}{{\mathcal{O}}} 

\usepackage{scalerel}
\usepackage{tikz}
\usetikzlibrary{svg.path}

\definecolor{orcidlogocol}{HTML}{A6CE39}
\tikzset{
  orcidlogo/.pic={
    \fill[orcidlogocol] svg{M256,128c0,70.7-57.3,128-128,128C57.3,256,0,198.7,0,128C0,57.3,57.3,0,128,0C198.7,0,256,57.3,256,128z};
    \fill[white] svg{M86.3,186.2H70.9V79.1h15.4v48.4V186.2z}
                 svg{M108.9,79.1h41.6c39.6,0,57,28.3,57,53.6c0,27.5-21.5,53.6-56.8,53.6h-41.8V79.1z M124.3,172.4h24.5c34.9,0,42.9-26.5,42.9-39.7c0-21.5-13.7-39.7-43.7-39.7h-23.7V172.4z}
                 svg{M88.7,56.8c0,5.5-4.5,10.1-10.1,10.1c-5.6,0-10.1-4.6-10.1-10.1c0-5.6,4.5-10.1,10.1-10.1C84.2,46.7,88.7,51.3,88.7,56.8z};
  }
}

\RequirePackage{etoolbox}
\DeclareRobustCommand\orcidicon[1]{\href{https://orcid.org/#1}{\mbox{\scalerel*{
\begin{tikzpicture}[yscale=-1,transform shape]
\pic{orcidlogo};
\end{tikzpicture}
}{|}}}} 
\usepackage{array}
\usepackage{makecell}

\newcommand{\type}{type\xspace}
\newcommand{\typearticle}{a\xspace}

\newcommand{\types}{types\xspace}

\newcommand{\topleveltype}{A_{\mathrm{Top}}}

\newcommand{\dltable}{T}

\newcommand{\SANTOS}{SANTOS\xspace}
\newcommand{\sFull}{$\mathrm{\SANTOS_{Full}}$\xspace}
\newcommand{\sKB}{$\mathrm{\SANTOS_{KB}}$\xspace}
\newcommand{\sSynth}{$\mathrm{\SANTOS_{Synth}}$\xspace}
\newcommand{\sCol}{$\mathrm{\SANTOS_{Col}}$\xspace}
\newcommand{\topk}{top-\textit{k}\xspace}
\newcommand{\Topk}{Top-\textit{k}\xspace}

\newcommand{\CSconf}{CS_{\textsc{Conf}}\xspace}
\newcommand{\RSconf}{RS_{\textsc{Conf}}\xspace}

\setcopyright{acmcopyright}
\copyrightyear{2022}
\acmYear{2022}
\acmDOI{}
\acmConference{}
\acmBooktitle{}
\acmPrice{15.00}
\acmISBN{978-1-4503-XXXX-X/18/06}
\begin{document}
\title{SANTOS: Relationship-based Semantic Table Union Search}

\author{Aamod Khatiwada*, Grace Fan*, Roee Shraga, Zixuan Chen, \\Wolfgang Gatterbauer, Ren\'ee J. Miller, Mirek Riedewald}
 \thanks{* Khatiwada and Fan are lead authors.}
\affiliation{%
  \institution{Northeastern University}
}
\email{ {khatiwada.a, fan.gr, r.shraga, chen.zixu, w.gatterbauer, miller, m.riedewald}@northeastern.edu}







\begin{abstract}
Existing techniques for unionable table search 
define unionability using metadata (tables must have the same or similar schemas) or column-based metrics (for example, the values in a table should be drawn from the same domain).  
In this work, we introduce the use of \emph{semantic relationships between pairs of columns} in a table to improve the accuracy of union search.  
Consequently, we introduce a new notion of unionability that considers relationships between columns, together with the semantics of columns, in a principled way. 
To do so, we
present two new methods to discover  semantic relationship  between pairs of columns.  
The first uses an existing knowledge base (KB), the second (which we call a ``synthesized KB'') 
uses knowledge from the data lake itself. 
We adopt an existing Table Union Search benchmark and present new (open) benchmarks that represent small and large real data lakes.
We show that our new unionability search algorithm, called \SANTOS,
outperforms a state-of-the-art union search that uses a wide variety of column-based semantics, including word embeddings and regular expressions. 
We show empirically 
that our synthesized KB improves the accuracy of union search by representing relationship semantics
that may not be contained in an available KB.
This result hints at a promising future of creating a synthesized KBs from data lakes with limited KB coverage and using them for union search.
\end{abstract}
\maketitle


The source code, data, and/or other artifacts have been made available at \url{https://github.com/northeastern-datalab/santos}.


\section{Introduction}\label{section:intro}

\begin{figure*}[ht]
  \includegraphics[scale = 0.53]{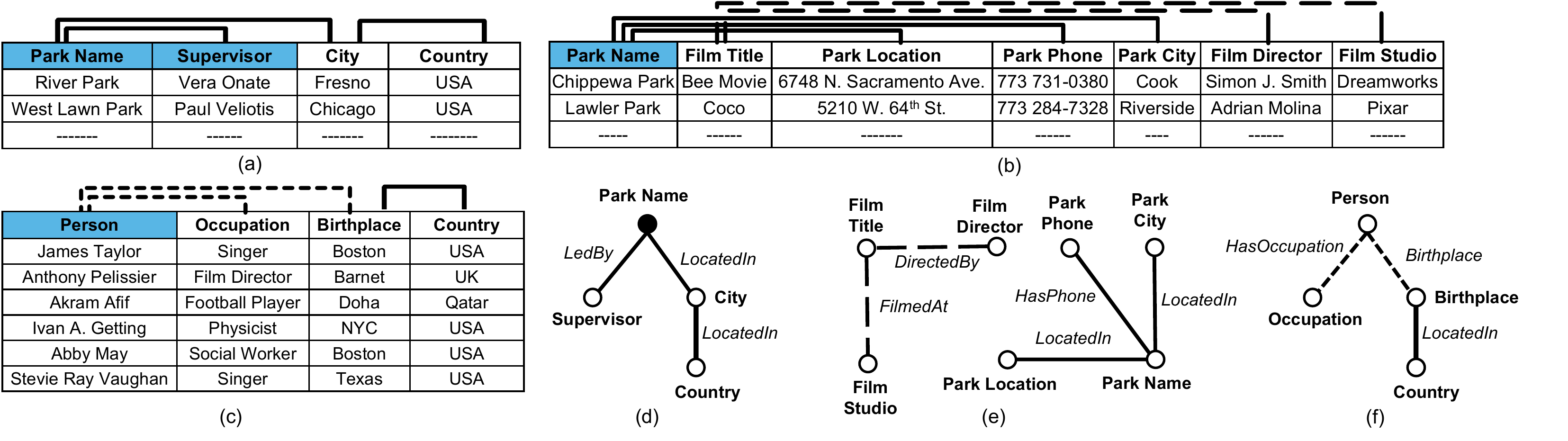}
  \caption{(a) 
  A table about parks. Relationships between \texttt{Park Name} and \texttt{Supervisor} (\texttt{ledBy}) and between \texttt{Park Name} and \texttt{City} (\texttt{locatedIn}) are found and depicted by the solid lines above the table.
 We have also found a relationship between  \texttt{Country} and \texttt{City} (\texttt{locatedIn}).
  (b) A 
  table 
  about parks and films shown in the park. Some of the relationships we found are shown by solid and dashed lines above the table.
  (c) A table about famous people.
  (d) A semantic graph for Table (a) with the relationships found related to the root \texttt{Park Name}.
    (e), (f) Semantic graphs for Tables (b), (c), respectively.
}
\label{fig:running_examples}
\end{figure*}

Table search is 
of growing practical relevance~\cite{2010_limaye_sunita_annotating_and_searching_web_tables,sarma_12_finding_related_tables,Zhu+16,DBLP:journals/debu/MillerNZCPA18}. 
It allows data scientists to find datasets 
needed for analysis or training of machine learning algorithms~\cite{nargesian_18_TUS, 2020_galhotra_s3d}.
Google's dataset search, for example, uses keyword search over metadata which 
Google incentivizes
data owners to use in a standard way~\cite{DBLP:conf/www/BrickleyBN19}.  
However, when dealing with enterprise data
or open data, 
keyword search via textual queries or over metadata is challenging
because metadata may be missing, inconsistent, or incomplete~\cite{DBLP:journals/pvldb/NargesianZMPA19, 2013_adelfio_schema_extraction, 2016_farid_clams, 2020_zhang_interactive_finding_tables}. 
Hence, table search for those domains must rely on the data available within given tables instead of relying on well-curated metadata~\cite{,2010_limaye_sunita_annotating_and_searching_web_tables,sarma_12_finding_related_tables}.

In this work, we
focus on 
\emph{table union search},
which is the problem of finding tables that can union with a query table and possibly extend it with more rows.
Prior work
by Nargesian et al.~\cite{nargesian_18_TUS} considers two tables to be \emph{unionable} if ``they share attributes from the same domain".  
We believe that this is a necessary, but not sufficient condition. 
In particular, we argue that the relationships modeled by pairs of attributes in a table should share a common semantic meaning.  We illustrate this with an example.

\begin{example}
\label{example:intent_column_importance}
Consider Tables (a), (b) and (c) in \cref{fig:running_examples} describing (a) parks, (b) films shown in different parks, and (c) the birth places of famous people, respectively. 
Suppose the user supplies Query Table (a)
and Tables (b), (c) reside in a data lake. The user wants to find data lake tables that union with 
the Query Table (a).  Typically, union search has been defined to permit either the full query table or a projection of it to union with the full data lake table or a projection of it~\cite{2020_bogatu_d3l}.
Now, if we only consider attribute unionability, Table (c) 
may be considered the most unionable table with three unionable attributes 
(\texttt{Person}, \texttt{Birthplace}, \texttt{Country})
having high attribute unionability scores with the attributes
(\texttt{Supervisor}, \texttt{City} and \texttt{Country}) 
in the query table.
This may be true if we are considering data values~\cite{nargesian_18_TUS} or if we consider the attribute names and values~\cite{DBLP:conf/ijcai/LingH0Y13,lehmberg_16_wdc_web_tables,2020_bogatu_d3l}.
However, if we consider the relationships in the tables, then we would notice that in Query Table (a), \texttt{City} is where the \texttt{Park} is located (and the person in \texttt{Supervisor} works), but in  Table (c),
\texttt{Birthplace} contains cities where the people in the \texttt{Person} column were born.
Deciding unionability based only on attribute unionability,
without considering the semantics of the relationships between attributes,
can lead to false positives.  Even worse, it can lead to the  omission of good answers like Table (b) which has information about parks and the city in which they are located, 
but might be considered less unionable than (c) based solely on column semantics.
\end{example}

We present a new definition of unionability based on both the attribute semantics and the 
\emph{relationship semantics between attributes}.
Relationships have been used to a limited extent
in web table search.  Their tables are often assumed to be ``entity tables'' 
(i.e., all attributes are properties of a known common entity represented by a single attribute)~\cite{DBLP:conf/semweb/MazumdarZ16, sarma_12_finding_related_tables,2020_deng_turl} or they use a subject column and a single relationship from that column to search for tables~\cite{venetis_11_recovering_semantics}.  
However, we do not assume that every table has a single subject attribute that all other attributes describe, 
nor that we can automatically detect such an attribute.
While 
the entity-table assumption 
may be mostly true in web tables, data lake tables (in both open data and enterprise lakes) tend to be much wider and may describe the interplay between more than one 
entity.

In this work, we provide a new definition of table union search.  
Given a query table $Q$ 
and a set of data lake tables $\mathcal{\dltable}$,
the \topk table union search problem is to find the best $k$ tables from $\mathcal{\dltable}$
that can be unioned with the query table by taking the column and relationship semantics of all involved tables into account.
\begin{example}
\label{example:relationship-unionability}
The columns \texttt{City} and \texttt{Country} in Table (a) are unionable with \texttt{Birthplace} and \texttt{Country} in Table (c) because the columns are unionable and their relationships are the same (\texttt{locatedIn}).  If a user is interested in amassing a collection of cities and their properties like location, then these tables are unionable.  However, if a user wants to collect a set of park supervisors and their properties (the city and country in which they work), then Tables (a) and (c) are not unionable.  This is because the relationship between \texttt{Supervisor} and \texttt{City} (\texttt{worksIn}) is not the same as the relationship between \texttt{Person} and \texttt{Birthplace} (\texttt{bornIn}).  A user may consider Table (b) to be unionable with Table (a), but only on $\pi$(\texttt{Park Name, City, Country}).  If we only consider attribute semantics, we might mistakenly say \texttt{Park Name, Supervisor, City} is unionable with \texttt{Park Name, Film Director, Park City}.  But the relationship semantics reveal that \texttt{Supervisor} has a different relationship to \texttt{Park Name} than \texttt{Film Director} has to \texttt{Park Name}.

\end{example}
\subsection{Contributions}

Our contributions can be summarized as follows.

\noindent $\bullet$
{\em Relationship-based Semantic Table Union Search}. We present a new definition of table union search that comprehensively integrates column and relationship semantics.\\
\noindent $\bullet$
{\em KB Solution}.  We present a solution to the relationship-based semantic table union search problem that uses a knowledge base (KB) to determine column and relationship semantics.  Our solution creates semantic graphs for tables whose nodes are columns and edges are binary relationships found in the KB (depicted in
\cref{fig:running_examples}(d)-(f) and explained in ~\cref{section:alt_relationship_unionability}).  We present a scoring function to match the semantic graph of a data lake table to that of a query table.\\
\noindent $\bullet$
{\em Synthesized KB Solution}.  We present a second, novel solution to the search problem that uses the data lake itself to determine if a set of columns have the same or similar semantics.  Our solution  determines if the binary relationship between two columns has similar semantics with other binary relationships in the data lake.  We show how this {\em synthesized KB} can be used to create semantic graphs and solve the table union search problem.  \\
\noindent $\bullet$
{\em Empirical Evaluation}. Our two solutions were designed to be used in concert with the Synthesized KB compensating for a curated KB that has only partial coverage over a data lake (which is typical when using both open KBs and proprietary enterprise KBs).  We show that when used together, our solutions outperform a state-of-the-art table union baseline.  We also experiment with using each solution independently.  Finally, we use the Synthesized KB together with different sized samples of the curated KB to understand how accurate our solution is as the  coverage of the curated KB decreases.\\
\noindent $\bullet$
{\em Benchmarks}. We develop (and openly share) three new relationship union benchmarks.  In the first, we reuse part of the TUS benchmark~\cite{nargesian_18_TUS} and label it with relationship semantics.  In the second, we use real open data tables and queries, and label a groundtruth.  The third is a larger data lake with open data tables and queries, where we only label the query results that either our technique or a baseline technique returns.  Hence, this larger benchmark can be used to evaluate precision and efficiency, but not recall. 

\section{Related Work}
\label{fig:related}
There is a rich literature on table union search that started with work on web tables.  
We begin by describing the state-of-the-art in table union search (\cref{section:related_search}), after which we discuss related work on attribute annotation (\cref{section:related_annotate}) and  other related work (\cref{section:related_others}).  

\subsection{Searching for Unionable Tables}\label{section:related_search}


\textbf{Cafarella et al.}~\cite{DBLP:journals/pvldb/CafarellaHK09} uses search-style keyword queries and returns a ranked list of relevant web tables. 
They mainly use tf-idf score and mean string length difference to find the similarity between the columns and combine them to infer the table similarity.

\textbf{Sarma et al.}~\cite{sarma_12_finding_related_tables} defines the problem of finding unionable web tables as an entity complement problem. 
Two tables are deemed unionable if they share similar schemas and a ``subject'' column, which is \emph{a single column that contains the entities the table is about}.
This work assumes that each table has a single subject column, a common trait of web tables, 
but as we argued in \cref{section:intro} is 
a limiting assumption in data lakes and open data.
Also, this approach relies solely on an existing KB. 

\textbf{Lehmberg et al.}~\cite{2017_lehmberg_stitching_web_tables} 
uses attribute labels and value overlap between attributes to determine table matching. 
They build on work from Ling et al.~\cite{DBLP:conf/ijcai/LingH0Y13} that relies on web tables having identical or similar schemas and uses value overlap.
However, attribute labels can be ambiguous or missing 
in  data tables~\cite{DBLP:journals/pvldb/NargesianZMPA19}, so we cannot rely on such metadata to be complete and consistent.

\textbf{Nargesian et al.}~\cite{nargesian_18_TUS} considers two tables to be unionable if they have a bipartite matching between a subset of their columns. 
This approach generates attribute unionability scores using three different statistical tests, one of which is semantic unionability that makes use of an existing KB.
Since they also look at set overlap and word embeddings, they do not rely exclusively on a KB.
However, they do not consider 
the semantics of relationships between columns as discussed in \cref{example:intent_column_importance}.
This in turn, may yield false positive results when searching for unionable tables.

\textbf{Fernandez et al.}~\cite{Fernandez2018SeepingSL} introduces SemProp, which links related tables by creating DAGs of table elements and external sources such as ontologies and embeddings, connected by semantic and syntactic links. Although the goals are similar, we solely leverage the data values in tables whereas SemProp mainly uses the tables' schemas and names to find related tables.

\textbf{Bogatu et al.}~\cite{2020_bogatu_d3l} proposes an 
attribute-unionability
framework similar to Nargesian et al.~\cite{nargesian_18_TUS} that adds attribute name similarity, 
regular expression similarity, and distribution similarity to determine the relatedness between 
tables.
We focus on attribute values, and also include relationship and type hierarchy semantics.

\subsection{Annotating Attributes and Relationships}\label{section:related_annotate}


\introparagraph{Supervised Approaches} 
Sherlock is a supervised technique to annotate attributes with 78 semantic types~\cite{hulsebos_19_sherlock}.
Zhang et al.~extended Sherlock with SATO, a hybrid machine learning model that uses topic modeling and structured learning~\cite{zhang_20_sato}
Recently, Suhara et al. introduced a multi-task learning approach called \textsc{Doduo} \cite{suhara_2021_doduo} that fine-tunes a BERT-base language model \cite{devlin_2019_bert} and predicts the column types and the relationship types between columns using column values.
These approaches, along with other supervised approaches~\cite{2010_limaye_sunita_annotating_and_searching_web_tables, DBLP:conf/aaai/TakeokaONO19} require training over annotated data. 
Due to the diversity of  data lakes (and the large number of possible column and relationship types) and the lack of comprehensive training data, it is difficult to employ a supervised approach 
to solve the table union problem.

\introparagraph{Unsupervised Approaches} There are several unsupervised approaches that annotate columns using KB's.
Mulwad et al.~\cite{2010_mulwad_Using_linked_data_to_interpret_tables} and Syed et al.~\cite{Exploiting_a_Web_of_Semantic_Data_for_Interpreting_Tables} base their work 
on KBs to find type labels for column headers.
Mazumdar and Zhang propose TableMiner+ \cite{zhang_17_tableminer+, DBLP:conf/semweb/MazumdarZ16}, which detects a subject column and incrementally finds attribute values in a KB to classify the columns. 
Venetis et al.~\cite{venetis_11_recovering_semantics} extracts data from the web and constructs an isA database for types and a relation database for properties (relationships) to provide a wide coverage of types and relationships.
These existing unsupervised approaches rely on information that is sparse in a data lake, such as column headers.
Our approach uses both an existing and a synthesized KB to find column and relation semantics from only attribute values. 
Also, we make  use of a KB type hierarchy to expand the range of union search.

\subsection{Other Related Work}\label{section:related_others}


\introparagraph{Embeddings} 
Nargesian et al.~\cite{nargesian_18_TUS} introduced the used of word embeddings for table union search.
Cappuzzo et al.~\cite{2020_cappuzzo_embeddings_for_relational_datasets} proposed to create relational embeddings for each table based on neighboring values in rows and columns.
They consider a few data integration tasks, but not unionability and require significant training data.










\introparagraph{Domain Discovery} Ota et al.~\cite{2020_ota_d4} introduce D4, an unsupervised algorithm that uses co-occurrence of values in different tables to discover the set of values that belong to the same domain. 
Our approach also makes use of co-occurrence of values when creating our synthesized KB, and 
we not only use co-occurrence of values in a column, but also consider the co-occurrence of values in binary relationships 
to discover relationship domains.

\introparagraph{Semantic Search over Structured Data} Galhotra et al.\ \cite{2020_galhotra_s3d} proposes S3D, a system that finds related tables, rows, and columns from KBs and datasets. 
S3D's 
search method is
similar to ours,
but they do not account for low KB coverage. 

\section{Relationship Unionability}
\label{section:alt_relationship_unionability}
We now introduce the building blocks of \SANTOS (SemANtic Table uniOn Search).
\SANTOS determines unionability based on not just the semantics of
columns (\emph{column semantics}),
but also the semantics of binary relationships between
columns (\emph{relationship semantics}).

\introparagraph{Column Semantics}
Like previous approaches to semantic annotation or column-type discovery in data lakes~\cite{2020_ota_d4,hulsebos_19_sherlock,zhang_20_sato}, 
we associate each column with a set of semantic annotations.
An annotation may be a type in a KB or, when using unsupervised techniques~\cite{nargesian_18_TUS,2020_ota_d4}, we may determine that a set of columns have the same, but unknown semantics (for example, they are \emph{attribute unionable}~\cite{nargesian_18_TUS}).   We call this \emph{column semantics}.
As an example, in Tables (a) and (b) of
\cref{fig:running_examples}, even if no known semantics can be found for Park information, 
we can still determine that the first columns 
share the same (but unknown) semantics.  
As the discovery process is uncertain, each 
annotation or type in the column semantics is associated with a confidence score. 

\begin{definition}[Column Semantics]
\label{def:cs}
Each column $c$ in a table $\dltable$ has a \emph{column semantics} (denoted $CS(c)$) which is a set of annotations each defining a conceptual domain to which the values in the column may belong.   
Each annotation $a \in CS(c)$ has a  confidence score between  0 and 1 that reflects the confidence in the inclusion of the annotation $a$ in $CS(c)$.
\end{definition}




\begin{example}
\label{ex:columnsemantics}
In Table (b) of \cref{fig:running_examples}, CS(\texttt{Park Name}) may have a set of types \texttt{\{Place, Tourist Attraction, Park\}} and CS(\texttt{Film Title}) may have \texttt{\{Creative Work, Movie\}}.  The confidence score will reflect our confidence in each type, for example if, hypothetically, more values in column \texttt{Park Name} map to type \texttt{Tourist Attraction} and only a few to the more specialized type \texttt{Park}, then a method may assign a lower confidence score to \texttt{Park}.  There may be other columns for which the column semantics is the empty set.
\end{example}



\introparagraph{Relationship Semantics}
In addition to column semantics used in previous table union search approaches~\cite{nargesian_18_TUS,2020_bogatu_d3l}, we use relationship semantics to guide the search.
Note that a relationship is any binary relation between column values (in particular, it does not need to be a function).
In \SANTOS,
each pair of columns is associated with a set of relationships, each with a confidence score.
Using a KB, we may assign a column pair with a known relationship (for example, a property) in the KB~\cite{tanon_20_yago, 21_hogan_knoweledge_graph_survery}, or we may use unsupervised techniques to determine that a set of column pairs have the same (but unknown) relationship.

\begin{definition}[Relationship Semantics]
\label{def:rs}

Each pair of columns $c_1$, $c_2$ in a table $\dltable$ has a \emph{relationship semantics} (denoted $RS(c_1, c_2)$) which is a set of annotations, each defining a conceptual relationship to which the tuples in $\pi_{c_1,c_2}(T)$  may belong.   
Each  annotation $a \in RS(c_1,c_2)$ has a  confidence score between  0 and 1 that reflects the confidence in the inclusion of the annotation $a$ in $RS(c_1,c_2)$. 
\end{definition}

\begin{example}
\label{ex:relationshipsemantics}
Relationship semantics in Table (b) of \cref{fig:running_examples} include \texttt{DirectedBy} between 
\texttt{Film Title} and \texttt{Film Director},
and \texttt{HasPhone} between \texttt{Park Name} and \texttt{Park Phone}.  
There may be no relationship discovered between \texttt{Film Title} and \texttt{Park Name}, 
in which case $RS(\texttt{Film Title, Park Name}) = \emptyset$.  
\end{example}

\introparagraph{Semantic Graph}
A relationship can be represented as 
a 
(Subject, Predicate, Object) triple 
or 
an edge in a graph whose  nodes  are
the 
columns of a table and edges connect pairs of 
columns.
For now, we assume we have an oracle (which may be a KB or
any relationship discovery tool) that provides column and relationship semantics.  We can form a \emph{semantic graph} for each table $\dltable$ that contains a node for each column and an edge between two columns if they have non-empty relationship semantics.


\begin{definition}[Semantic Graph]
\label{def:semanticgraph}
Given a set of columns $\mathcal C = \{c_1, c_2 \dots c_m\}$ in a table $\dltable$, 
the semantic graph of  $\dltable$ 
is a graph 
$SG(\dltable) = (V,E)$
with a distinct vertex $v_i$ for each column $c_i$, labeled with $CS(c_i)$.
For each column pair $(c_i, c_j), i \neq j$, with non-empty relationship semantics $RS(c_i, c_j) \neq \emptyset$, there is
an undirected edge $e_{ij}$ between $v_i$ and $v_j$
labeled with $RS(c_i, c_j)$.
\end{definition}




\begin{example}
\label{ex:semanticgraph}
Semantic graphs for data lake Table (b) about parks and films and Table (c) about people are shown in \cref{fig:running_examples} (e), (f) respectively (with column semantics omitted). As seen in (e),  semantic graphs of data lake tables may not be connected, depending on the relationship discovery. 
Some edges are dashed only to make the connection with Tables (b) and (c) more intuitive.
\end{example}

\introparagraph{Unionability Search over Semantic Graphs}
In \SANTOS, a user (a data scientist) provides a Query Table (denoted by $Q$) and specifies its \textit{intent column} (denoted by $I$), which is a column of most interest to the user that forms relationships of interest with other columns.
\SANTOS creates a semantic graph for the query table that is restricted to being a tree rooted at the intent column, called the \emph{Query Semantic Tree}. We find relationships from the intent column to other columns in the query table, then transitively search for relationships from these columns.
This process allows a user to direct the search to include their
intent column.
We search for data lake tables that contain a column with a similar (matching) semantics to the intent column and with similar relationships.
More specifically, we will look for a tree within each semantic graph of data lake tables that matches a subtree of the query tree rooted at the intent column.
We do not require that the full query semantic tree be covered by a data lake table.  We assume a scoring function that considers the strength of column and relationship matching as well as how much of the query tree is matched.
We define a precise scoring function in \cref{section:semantic_union_search}, after we describe how we compute column and relationship semantics. The scoring function captures how closely the Semantic Graph of a data lake table matches with the Query Semantic Tree.

\begin{example}
\label{example:definition_motivation}
Consider Table (a) about parks in
\cref{fig:running_examples} as a query table. Suppose a user's intent column is
\texttt{Park Name}, which has two relationships  
(1) \texttt{Park Name--LedBy--Supervisor} and 
(2) \texttt{Park Name--LocatedIn--City}.
\texttt{City} forms a third relationship:
\texttt{City--LocatedIn--Country}.
The semantic graph is depicted in \cref{fig:running_examples}~(d). For brevity, we omit the CS in the figure and include only one RS for each edge.
With the intent column I, 
we can determine if the semantic graph of Table (b) (\cref{fig:running_examples}(e)) contains a subtree that maps to the query semantic tree rooted at I.  If the CS for 
\texttt{Park Name} and \texttt{City} match in (a) and (b) and RS of \texttt{locatedIn} matches \texttt{locatedIn},
then (b) is a candidate unionable table that unions on $\Pi_{\texttt{Park Name, Park City}}$.
Notice however the semantic graph \cref{fig:running_examples}(f) of \cref{fig:running_examples}(c)
about people.
Although there is a possible matching relationship involving \texttt{Birthplace} and \texttt{City} (with \texttt{City--LocatedIn--Country}), (c) is not a candidate unionable table as there is no good match with the user's intent column.
\end{example}
\begin{definition}[\SANTOS \Topk Union Search Solution]
\label{def:new_santos_unionability}
Given a set of Data Lake Tables $\mathcal{\dltable}$, a query table $Q$ with a specified intent column $I$, 
a semantic graph for $Q$, $SG(Q)$, which forms a tree rooted at $I$.
We assume a 
semantic graph matching or 
scoring function $S$ that for each $\dltable \in \mathcal{\dltable}$ returns the highest scoring subtree of $SG(\dltable)$ that matches a subtree of $SG(Q)$ rooted at $I$ along with its score.
The \SANTOS union search solution is the \topk data lake tables having the highest score modeled by $S$.
\end{definition}
We introduce two methods to create semantic graphs.
 \cref{section:unionability_components}
describes one that uses an existing KB, and \cref{section:synthetic_knowledge_base} introduces another that is designed to be used when there is no or partial KB coverage over the data lake
(as commonly seen with enterprise and open data lakes). We discuss how we leverage these methods to create semantic graphs in \cref{section:semantic_union_search} and walk through our implementation in \cref{section:santos_implementation}. Finally, we
present experiments (\cref{section:experiments})  using each method independently and together to provide better accuracy.

\section{Semantic Graph Creation with KB}
\label{section:unionability_components}


In this section, we present a method for creating semantic graphs using an existing KB. 
Recall that metadata (like column headers, table name, etc.) may be missing, inconsistent or incomplete in data lake tables \cite{DBLP:journals/pvldb/NargesianZMPA19, 2013_adelfio_schema_extraction, 2016_farid_clams, 2020_zhang_interactive_finding_tables}. 
Therefore, we create the semantic graphs by using the cell values only.

\subsection{Column Semantics}


\label{section:column_semantics}
To create column semantics (CS), we associate a set of KB \types\  (called annotations in \cref{def:cs}) to each column.
Associating columns with \types from a KB is a 
well-studied problem~\cite{venetis_11_recovering_semantics,ritze_15_matching_tables_to_dbpedia,zhang_17_tableminer+,hulsebos_19_sherlock,zhang_20_sato}.
Like previous work on unionability~\cite{nargesian_18_TUS} and 
column \type detection~\cite{zhang_20_sato,2020_ota_d4, suhara_2021_doduo}, we only use values within a column to determine the associated KB \types.
Like others~\cite{nargesian_18_TUS, venetis_11_recovering_semantics}, we use both the KB types and type hierarchy to define CS, since 
there may be tables that match with a query table at a different granular level. Therefore, instead of annotating each column with a single type, we annotate with a set of types. This provides flexibility in matching a data lake table with query tables of different granularity.

\begin{example}
\label{example:why_hierarchy}
Consider the \texttt{Birthplace} column in 
Table(c) of \cref{fig:running_examples} 
that describes where people were born.
We might assign a more specific type \texttt{city} to this column because the majority of values are cities. 
However, we might also assign a broader type \texttt{Place} as this column also contains information on places that are not cities (e.g., \texttt{Texas} and \texttt{Barnet}). 
Consequently, using any single \type  for the columns in the data lake tables
can impact the effectiveness of union search as it may differ from the detected \type in the query table.
Therefore, we keep a set of \types as
the CS
and select one 
to use at query time depending on the query table CS. This allows us to match the same data lake table column with the columns from different query tables--some with just cities and some with places. 
\end{example}
CS is assigned based on \emph{semantic consistency}.
For example, if a column is assigned \typearticle \type \textit{place}, it can also be assigned another granular \type \textit{city} but not \textit{music album}, which is semantically inconsistent. We make use of the KB type hierarchy to ensure semantic consistency by only assigning types to a column that are 
in a ISA relationship in the KB type hierarchy.
\SANTOS can be used with any 
open, enterprise-level, or domain-specific KB.
We use YAGO 4~\cite{tanon_20_yago} (referred to as YAGO hereafter) in our setup which has a single root.
The KB root is a generic type and is uninformative to use as part of CS.
As YAGO has a large and rich set of types that are direct descendants of the root, we choose to use all direct descendants of this root as the top level types denoted by $\topleveltype$. 
These are \types that can form the root of CS and are assumed to be semantically disjoint.\footnote{In a different KB, a different choice may be made.}
%
To define the CS($c$), we map each value in $c$ to the KB. If the value appears in the KB in a type $a$, we add $a$ and all its ancestors up to a top level type $\in \topleveltype$ to the column semantics candidate set $CS_{\mathrm{candidate}}(c)$. 
Note that a single value may map to multiple types (for example in YAGO, \texttt{Boston}\footnote{\url{https://en.wikipedia.org/wiki/Boston_(album)}} maps to \texttt{City} and
\texttt{Music Album}) and different values may map to different or identical
types. 
After going through all values, if $CS_{\mathrm{candidate}}(c)$ contains multiple top level types, then we keep only the top level type 
(and all its descendants) in $CS_{\mathrm{candidate}}(c)$ mapped to by the majority of values, which ensures semantic consistency.\footnote{In the rare case of ties, we keep the rarer top level type having the fewest entities.
} 

\begin{example}
\label{example:top_type_explanation }
Consider \texttt{Birthplace} column in Table (c) of \cref{fig:running_examples} that has 5 unique values.
The value \texttt{Boston} 
is associated with \texttt{place} and \texttt{creative work}, both of which are in $\topleveltype$. 
The KB returns both types for the {\tt Birthplace} column, as well as descendants of \texttt{place} to which at least one value is mapped: \texttt{\{administrative area, city, state\}} and the descendant of \texttt{creative work} to which at least one value is mapped: \texttt{\{music album\}}.
Notice the majority of the values 
are associated with \texttt{place}. 
Therefore, we select \texttt{place} as the top level type for the {\tt Birthplace} column and discard \texttt{creative work} and its descendants.
Therefore, CS(\texttt{Birthplace}) is the set of types:
\texttt{\{place, administrative area, city, state\}}.
\end{example}




\subsection{Column Semantics Confidence}
\label{subsection:csconfidence}




We assign a confidence score to each \type by taking the product of the
\emph{frequency score} ($\mathit{fs}$) and \emph{granularity score} ($\mathit{gs}$). 
This score captures the same intuition as the well known measure TF-IDF ~\cite{aizawa2003information, ramos2003using}. 
TF-IDF measures the importance of a word in a document considering its occurrence and specificity. Here we want to measure the importance of each type. 
In our case, TF, 
which is the occurrence of a term in the document~\cite{1957_luhn_tf}, aligns with $\mathit{fs}$ and IDF, 
which captures the specificity and rareness of the term~\cite{1988_sparck_idf}, aligns with $\mathit{gs}$.

Concretely, the \textit{frequency score} $\mathit{fs(a)}$ of type $a \in CS(c)$ 
in column $c$
is the fraction of unique values in $c$ that are mapped to $a$ ($|c_a|$) out of all unique values in $c$ mapped to the KB ($|c_{\text{KB}}|$):
\begin{equation}
    \mathit{fs(a)} = \frac{|c_a|}{|c_{\text{KB}}|}
\end{equation}

\begin{example}
\label{example:confidence_explanation }
Consider the CS 
for the column {\tt Birthplace}  of~\cref{fig:running_examples}(c) which includes the annotations \texttt{city} and \texttt{state}.
These annotations are siblings in the KB, each with an ISA relationship to a parent \texttt{place} 
Notice however, that only 1 unique data value (\texttt{Texas}) out of 5 is associated with \texttt{state} while 3 are associated with \texttt{city} (\texttt{Barnet} is mapped to neither \texttt{city} nor \texttt{state} but does map to \texttt{place}). 
To model this difference, we assign frequency scores fs(\texttt{state}) = 0.2 and fs(\texttt{city}) = 0.6. 
As all values are mapped to the top level type \texttt{place}, fs(\texttt{place}) = 1.0.
\end{example}

A set of entities of a descendant type is always a subset of entities belonging to its ancestor types in the type hierarchy. So, the
higher level types are always mapped to by the same or more entities than children types.  This is reflected in the frequency score.
From the perspective of information theory, the less probable outcome has greater information \cite{mackay2003information, 1988_sparck_idf}.
In our case, a type that appears more frequently in the KB is less informative.
As mentioned earlier, inspired by the well-known tf–idf measure~\cite{aizawa2003information}, we penalize the frequent ancestor types by using a frequency-based strategy utilizing KB statistics and assign a \textit{granularity score (gs)} to each type.
The basic idea is to count the frequency of each type $a$ by counting the entities that map to type $a$ and penalize the frequent \types.
The granularity score is computed in \cref{eq:type_score} which uses a log function. 

\begin{equation}\label{eq:type_score}
    \mathit{gs(a)} = \frac{1}{min(1, \log(a.count))}
\end{equation}
To keep $\mathit{gs}$ consistent with $\mathit{fs}$, $\mathit{gs}$ also ranges from 0 to 1. 
Note that for rare types with less than 10 entities, the log function returns a value less than 1. Thus, we use the \emph{min} function in the denominator.



\begin{example}
Consider the entity \texttt{Boston}, which belongs to the \types \texttt{city} and \texttt{place}.
In YAGO, 
over 6 million entities have \type \texttt{place}, and $\sim$42,000 entities have \type \texttt{city}. 
Then, \texttt{city} is a more informative \type than \texttt{place}, which is an ancestor \type.
Hence, the granularity scores are $gs(\texttt{place}) \approx 0.14$ and $gs(\texttt{city}) \approx 0.22$.
\end{example}

We assume that the computations of \emph{granularity score} and \emph{frequency score} are independent of one another as the former is based on KB statistics and the latter is based on the semantics of values in a column.
Our objective is to prioritize those \types that are more specific (i.e. higher \emph{granularity score}) and also capture the semantics of the column values better (i.e. high  \emph{frequency score}). 
Thus, given a column $c$ with annotation $a$, we define the KB Column Semantics Confidence Score by aggregating $\mathit{fs}$ 
and $\mathit{gs}$ of $a$:
\begin{equation}
\label{eq:column_kb_score}
    \CSconf(c, a) = 
    \begin{cases}
          \mathit{fs(a)} \cdot \mathit{gs(a)} \quad &\text{if} \, c\in\text{data-lake table } T \\
          \textit{fs(a)} \quad &\text{if} \, c\in\text{query table } Q
     \end{cases}
\end{equation}


Note that to avoid double penalization, 
we only penalize the top-level types in the data-lake tables.  


\begin{example}
\label{example:column_semantics}
\texttt{Birthplace} in data lake table \cref{fig:running_examples}(c) has $CS$ with respective $\CSconf$
\texttt{\{place:$1.0\cdot0.14$, 
administrative area:$1.0\cdot0.17$, 
city:$0.60\cdot0.22$, state:$0.20\cdot0.35$\}}.
\end{example}

\subsection{Relationship Semantics}
\label{subsection:relation_semantics}

We compute relationship semantics (RS) for every 
pair of string (non-numeric) columns within the query table and within  data lake tables.
Note that many pairs of columns (like {\tt Park Name} and {\tt Film Director}) may not have a semantic relationship represented in the KB.  Intuitively, there is at least an indirect relationship between all columns in a table (in this case, the director of a film shown in the park), but in this section, we are only interested in relationships found in the KB.
Suppose columns $c_1$ and $c_2$ in table $\dltable$ have non-empty CS (meaning they are annotated with at least one KB type). 
To determine RS, we determine if a pair of values in $\pi_{c_1,c_2}(\dltable)$ 
is associated with entities $(e_1,e_2)$ that have a KB relationship $p$.
The RS Confidence Score $\RSconf(c_i,p,c_j)$ for a binary relationship $p$ between columns $c_i,c_j$ ($\pi_{c_i, c_j}(T)$) in $T$ is:
\begin{equation}
    \RSconf(c_i,p, c_j) = \frac{|(c_i, c_j)_p|}{|(c_i, c_j)_{\text{KB}}|}
\end{equation}
such that $|(c_i, c_j)_p|$ is the number of unique value-pairs with predicate $p$ from KB, and $|(c_i, c_j)_{\text{KB}}|$ is the total number of unique value pairs mapped to the KB. Note that only the relationship semantics with the maximum score is included in the semantic graph.\footnote{If there is a tie, we pick the annotation for $p$ that has the smallest number of entity pairs in the KB, thus preferring the rarer predicate.} 


\begin{example}
\label{example:rscs}
In \cref{fig:running_examples}(c),
RS(\texttt{Person,Birthplace}) contains the annotation \texttt{birthplace} with confidence score
$\RSconf$ = 1.0.
\end{example}
\section{Synthesized KB Semantic Graph}
\label{section:synthetic_knowledge_base}
\begin{figure*}[ht]
  \includegraphics[scale = 0.52]{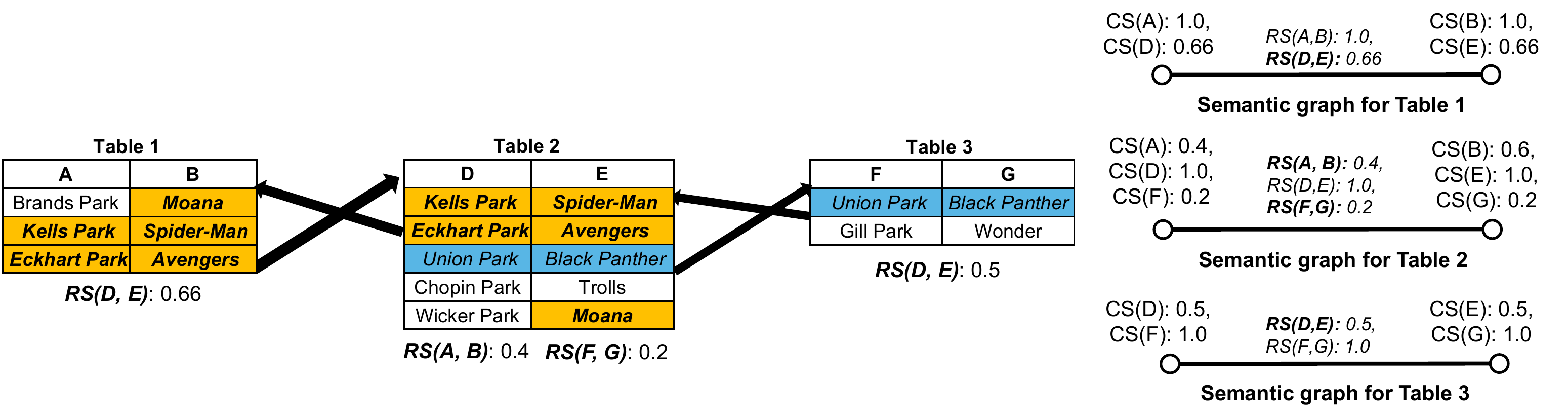}
  \caption{Synthesized relationship semantics of data lake tables and their respective semantic graphs. Value-pairs bolded and highlighted in orange appear in Tables 1 and 2, and the value-pair italicized and highlighted in blue appear in Tables 2 and 3.
  }
\label{fig:synthetic_kb}
\end{figure*}

KBs may have limited coverage over 
real data lakes. 
Hence, using only an
existing
KB (even a set of KBs) 
to determine CS and RS can lead to low coverage.
Our experimental study 
indicates that YAGO~\cite{tanon_20_yago}, a well-known and maintained KB, covers only $42\%$ 
of the string cell values in UK open data and $34\%$ in Canada open data. 
To solve this problem, we propose a novel
data-driven approach using the data lake itself, creating a \emph{synthesized KB}. 
Our key insight is that we can 
\emph{replace the role of an existing KB in finding CS and RS with a KB that captures co-occurrence information across data lake tables}.
To determine CS, instead of mapping values to an existing KB, we now use a mapping to other columns with the same values.
Rather than finding actual semantic types, we leverage \type co-occurrence across 
columns to decide their semantic compatibility.
To do so, we annotate all values that co-occur in a column or a meaningful relationship with a ``synthesized type.''
We then determine CS and RS by also considering other column and column pairs (thus their synthesized types) with overlapping values.


\subsection{Synthesized Column Semantics}
\label{subsection:theory_synthesized_cs}
Generally, values within the same column share the same semantic types.
For example, in \cref{fig:running_examples}(c), all values in the \texttt{Person} column 
have \type \texttt{person} and values in \texttt{Birthplace} share \type \texttt{place}. 
Considering this property and assuming we do not have homographs \cite{21_leventidis_homograph_detection}, 
we start by assigning to each column in the data lake a unique synthesized column semantics with a confidence score equal to 1.
For example, considering Table 1 about parks and movies in \cref{fig:synthetic_kb}, we assign synthesized column semantics \texttt{A}, \texttt{B} to the columns. Similarly, we assign \texttt{D}, \texttt{E} and \texttt{F}, \texttt{G} as column semantics to the columns in Tables 2, 3 respectively. 
There can also be different columns 
that share the same column semantics.
For instance, consider Tables 1 and 2 in \cref{fig:synthetic_kb}. 
Notice that columns \texttt{A} and \texttt{D} are both about parks while columns \texttt{B} and \texttt{E} are about movies. 
We hypothesize that columns with common semantics have overlapping values that share the same types. 
We want to determine how likely it is for a column $c$ to share semantics with column $c_j$, and thus inherit $CS(c_j)$. 
Formally, along with its own column semantics,
Like in \cref{subsection:csconfidence}, we also assign synthesized \type $a$ from column $c_j$ to column $c$ with a confidence score. 
We define a synthesized column semantics confidence score for column $c$ where $\CSconf$ of $a$ is the fraction of unique values in $c$ that are also in $c_j$ ($|c \cap c_j|$) over the total unique values in $c$ ($|c|$). If $c = c_j$, then $\CSconf$ of $a$ is 1.
\begin{equation}\label{eq:column_synth_score}
    \CSconf(c, a \in CS(c_j)) =
    \begin{cases}
            1 \quad &\text{if} \, c = c_j\\
          \dfrac{|c \cap c_{j}|}{|c|} \quad &\text{otherwise}
     \end{cases}
\end{equation}
Note that as we do not have type hierarchy information for the synthesized column semantics, we assume all synthesized CS are of the same granularity level and set $gs(a) = 1$ for each \type.

\subsection{Synthesized Relationship Semantics}
\label{subsection:theory_synth_relationship}
For column pairs in a table that share a \emph{"meaningful relationship"}, 
we assign a synthesized relationship semantics (RS). For example, consider Table 1 of \cref{fig:synthetic_kb} about films shown at parks, where columns \texttt{A} and \texttt{B} have a meaningful relationship, annotated as $RS(A,B)$.
Following the same logic as annotating synthesized CS, we consider two column pairs to have the same RS if their value pairs overlap.
For instance, column pairs (\texttt{A, B}) and (\texttt{D, E}) in Table 2 of \cref{fig:synthetic_kb} have overlapping value pairs (depicted in bold text), indicating that they likely share a meaningful relationship.
Therefore, along with its own synthesized RS, we also assign a synthesized RS $p \in RS(d_i, d_j)$ 
of column pair $(d_i, d_j)$ to the column pair $(c_i, c_j)$ ($\pi_{c_i, c_j}(T)$) of table $T$ with synthesized $\RSconf$ given by:
\begin{equation}\label{eq:relation_synth_score}
    \RSconf(c_i, p, c_j) = 
    \begin{cases}
        1 \quad &\text{if} \, c_i = d_i, c_j = d_j\\
       \dfrac{|(c_i, c_j)\cap(d_i,d_j)|}{|(c_i, c_j)|} \quad &\text{otherwise}
 \end{cases}
\end{equation}
where $|(c_i, c_j)\cap(d_i,d_j)|$ is the number of unique overlapping value pairs between distinct column pairs and $|(c_i, c_j)|$ is the total unique value-pairs in $(c_i, c_j)$. 

Note that although we may have multiple types in relationship semantics for each column pair (seen in Semantic Graphs in \cref{fig:synthetic_kb}), we use only the \type that best matches
with the query table during the query phase (\cref{section:semantic_union_search}). 
We discuss implementation details for efficient indexing of the synthesized KB in \cref{section:santos_implementation}.



\section{\SANTOS Union Search}
\label{section:semantic_union_search}
%

In the previous sections, we presented two methods for creating
semantic graphs: one relying on a high-quality existing KB (YAGO), the other on our
proposed synthesized KB.
We now discuss how to compute a unionability score from both semantic graphs.

\introparagraph{Scoring function}
Recall that \SANTOS takes a query table as input from the user and the objective is to generate a ranked list of top-k unionable tables. Generating a ranked list of relevant documents for a given query is a well-studied problem in the literature~\cite{GOSWAMI2017rankingfunctions, frank2005comparingrankingfunctions}. Recently, Ho et al. formulated a ranking function to rank a list of relevant text documents for a given sentence query~\cite{2019_vinh_thinh_qsearch}. They extract a set of tokens representing the 
context of the query sentence and each candidate document. 
Each token is assigned 
a confidence score. The (normalized) summation of the confidence scores of matching tokens is used to rank the documents. 
This function is modified and used to rank the relevant web tables for the given quantity queries~\cite{2021_vinh_thinh_qute}.
We use the same concept to motivate our scoring function
without normalization as this does not change the ranking.
In our work,
the context is represented by matching query table's column pairs with data lake table's column pairs and the confidence is represented by their matching quality. Specifically, we match a connected subtree of $Q$'s semantic query tree, rooted at the intent column, into 
the semantic graph of data lake tables.
Intuitively, the larger the number of matching column pairs and the higher the
confidence that the corresponding semantic types agree, the higher the match
score.
Note that in the score computation, nodes and edges may each
be annotated with multiple possible CS and RS
(\cref{section:column_semantics}, \cref{subsection:theory_synth_relationship}),
each with respective confidence scores.

\begin{example}
\label{example:multiple_annotations}
Consider table (a) of \cref{fig:running_examples} about parks.
Possible semantics for column pair (\texttt{Park Name}, \texttt{City}) are
\texttt{park-locatedin-place}, \texttt{park-locatedin-city}, etc. based on the given KB.
When value pairs from these columns are not covered by the given KB, we also obtain semantics
from the synthesized KB such as \texttt{CS(W)-RS(W,X)-CS(X)} and
\texttt{CS(Y)-RS(Y,Z)-CS(Z)}, where \texttt{CS(W)}, \texttt{CS(X)}, \texttt{CS(Y)} and \texttt{CS(Z)} are the synthesized
CS and \texttt{RS(W,X)}, \texttt{RS(Y,Z)} the synthesized RS.
\end{example}

With semantics from both the KB and synthesized KB for the same column pair, we propose a 2-step
approach for computing the unionability score.
First, an ``intra-method'' technique selects the best semantic match between query
and data lake table for each source separately.
Then an ``inter-method'' comparison selects the semantic match that maximizes
the overall unionability score.

Let $a_1, a_2 \dots a_x = (CS(Q_c) \cap CS(\dltable_c))$ 
be the intersecting CS between a query table Q's column $Q_c$ and a data lake
table $\dltable$'s column $\dltable_c$ given by a semantic graph creation method $G$
(i.e., either KB or the synthesized KB).
For each $i$, let the corresponding confidence scores be
$\CSconf(Q_c,a_i)$ and $\CSconf(\dltable_c,a_i)$, respectively.
Since the column semantics assigned to the query table and data lake table are independent of each other, we take a product of confidence scores for each token to get their match score. Then, we select the match that maximizes the score.
Intuitively, the match score between $Q_c$ and $\dltable_c$ is determined by the column semantics
with the greatest product of confidence scores:
\begin{multline}
\label{eq:col_match}
    colMatch_G(Q_c, \dltable_c) =
    \max_i \; \CSconf(Q_c, a_i) \cdot \CSconf(\dltable_c, a_i).
\end{multline}

We determine the relationship match score for column pairs $(Q_{c1}, Q_{c2})$ in $Q$
and $(\dltable_{c1}, \dltable_{c2})$ in $\dltable$ analogously as:
\begin{multline}
\label{eq:rel_match}
    relMatch_G((Q_{c1}, Q_{c2}),(\dltable_{c1}, \dltable_{c2})) \\
    = \max_i \; \RSconf(Q_{c1}, p_i, Q_{c2}) \cdot \RSconf(\dltable_{c1}, p_i, \dltable_{c2}).
\end{multline}
Here $p_1, p_2,\ldots, p_x = (RS(Q_{c1}, Q_{c2}) \cap RS(\dltable_{c1}, \dltable_{c2}))$
are the intersecting relationship semantics between the column pairs.
Note that depending on the KB, $RS(T_{c1}, T_{c2})$ may differ from $RS(T_{c2}, T_{c1})$. Therefore, the KB may return $RS(T_{c1}, T_{c2})$ for the data lake table and $RS(Q_{c2}, Q_{c1})$ for the query table. To model this, we preserve both $RS(T_{c1}, T_{c2})$ and $RS(T_{c2}, T_{c1})$ for the data lake table. Once we get RS for the query table, we match the data lake table's RS that maximizes the score with respect to the query table's RS in \cref{eq:rel_match}.


The overall match score between $Q$'s column pair ($Q_{c1}, Q_{c2}$) and $\dltable$'s
column pair ($T_{c1}, T_{c2}$) based on method $G$ is then computed as:
\begin{multline}
\label{eq:pair_match_individual}
    pairMatch_G((Q_{c1}, Q_{c2}), (T_{c1}, T_{c2})) = colMatch_G(Q_{c1}, T_{c1}) \\
    \cdot relMatch_G((Q_{c1}, Q_{c2}), (T_{c1}, T_{c2}))
    \cdot colMatch_G(Q_{c2}, T_{c2})
\end{multline}

\begin{figure}[h]
    \centering
  \includegraphics[width=\linewidth]{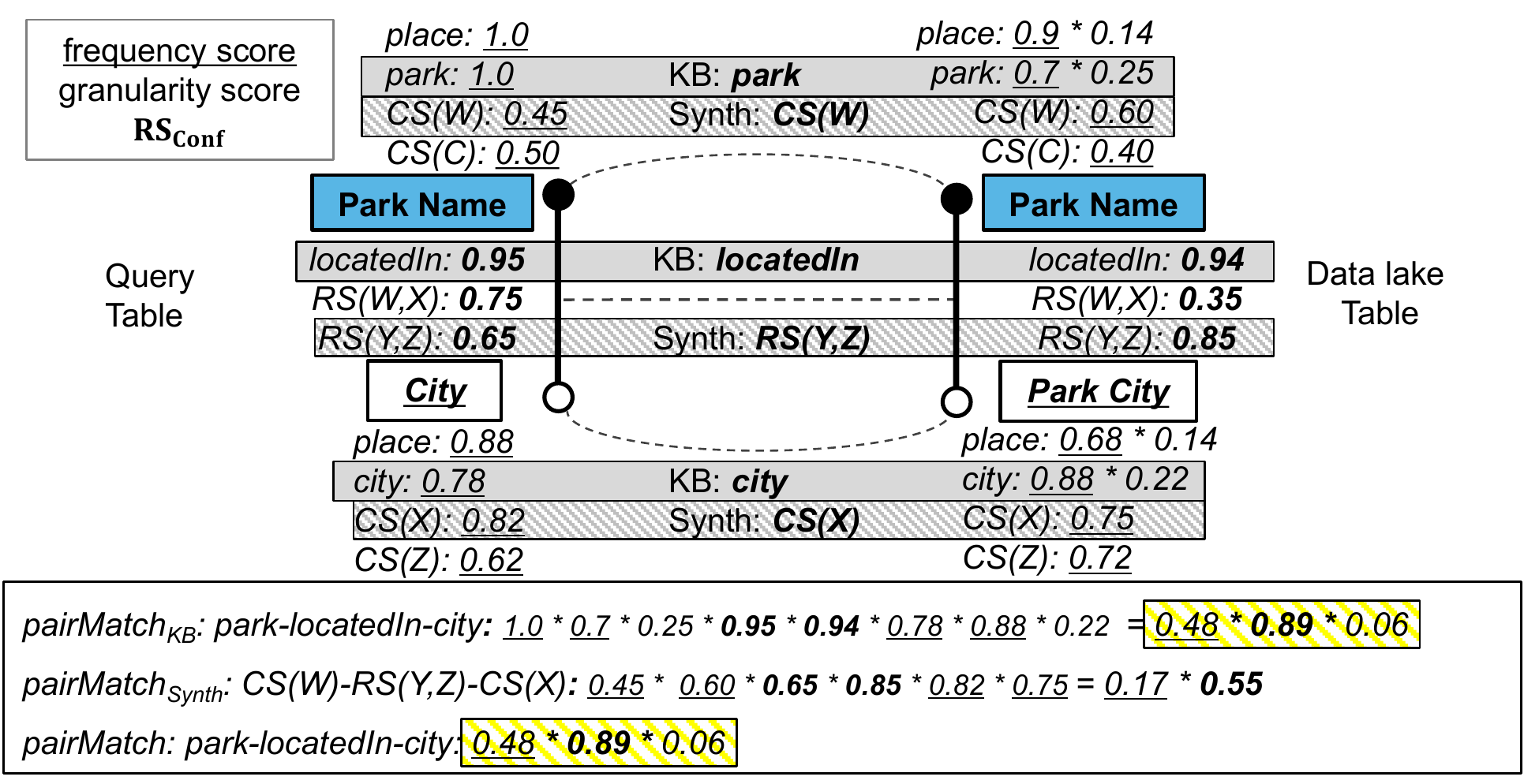}
  \caption{Computation of pairMatch score between the semantic tree of the parks table
  \cref{fig:running_examples}(a) as query table (left) and the parks-and-movies table
  \cref{fig:running_examples}(b) as data lake table (right). From the data lake table's
  semantic graph, we extract the semantic tree rooted at \texttt{Park Name}.
  The dotted arcs connect the matching columns and relationships.
}
\label{fig:score_calculation}
\end{figure}

Let $KB$ and $Synth$ represent the semantic graph creation methods based on the given
and synthesized KB, respectively. Recall that $\CSconf$ based on the given KB is assigned
a granularity score for penalizing the top-level types (\cref{eq:column_kb_score}).
However, due to the absence of a type hierarchy in the synthesized KB, we consider all
types to be of same level and assign a granularity score of 1 to each
synthesized column semantics (\cref{eq:column_synth_score}). 
To avoid bias, we compare the ``inter-method'' pair matches ignoring
the granularity score.

Formally, let $a_1$ and $a_2$ be the column semantics selected in
$pairMatch_{KB}((Q_{c1}, Q_{c2}), (T_{c1}, T_{c2}))$.
With $f$ denoting a flag such that $f = 1$ iff
$\frac{pairMatch_{KB}((Q_{c1}, Q_{c2}), (T_{c1}, T_{c2}))}{gs(a_1).gs(a_2)} \geq  pairMatch_{Synth}((Q_{c1}, Q_{c2}), (T_{c1}, T_{c2}))$, the match score between
($Q_{c1}, Q_{c2}$) and ($T_{c1}, T_{c2}$) is computed as:
\begin{multline}
\label{eq:pair_match}
    pairMatch((Q_{c1}, Q_{c2}), (T_{c1}, T_{c2})) = \\
    \begin{cases}
            pairMatch_{KB}((Q_{c1}, Q_{c2}), (T_{c1}, T_{c2})) \quad &\text{if} \, f = 1\\
            pairMatch_{Synth}((Q_{c1}, Q_{c2}), (T_{c1}, T_{c2})) \quad &\text{otherwise}
     \end{cases}
\end{multline}

\begin{example}
\label{example:unionability_score}
Let Table (a) in \cref{fig:running_examples} about parks be the query table ($Q$)
with \texttt{Park Name} as the intent column.
Consider Table (b) in \cref{fig:running_examples} about parks and movies as the data lake
table ($\dltable$). There is a possible match between these two tables such that
\texttt{Park Name} matches with \texttt{Park Name} and
\texttt{City} matches with \texttt{Park City}. We compute the corresponding pair match
score as follows.
\cref{fig:score_calculation} shows the semantic graphs and the scores involved in
selecting the matching pairs. The query table semantic tree is shown on the left and
the semantic tree extracted from the data lake semantic graph rooted at \texttt{Park Name}
is shown on the right. The semantic graph before extracting the tree is shown in
\cref{fig:running_examples}(e).
The dotted arcs connect the matching nodes and edges.
First, we use \cref{eq:col_match} to find the matching column semantics at each node
between the tables using existing and synthesized KBs.
For \texttt{Park Name} and \texttt{Park Name}, 
\texttt{park} and \texttt{CS(W)} are selected from the existing and synthesized KB, respectively.
Also for \texttt{City} and \texttt{Park City}, \texttt{city} and \texttt{CS(X)}
are selected using the same formula (\cref{eq:col_match}).
Furthermore, as we have only one relationship semantics for KB, i.e.,
\texttt{locatedIn}, it is selected as the winner.
For the synthesized KB method, \texttt{RS(Y,Z)} wins over \texttt{RS(W,X)}
according to \cref{eq:rel_match}.
Finally, we use \cref{eq:pair_match} to perform the inter-method comparison between
\texttt{park--locatedIn--city} and \texttt{CS(W)--RS(Y,Z)--CS(X)}.
As shown in the figure, as $max(0.48 \cdot 0.893, 0.166 \cdot 0.552) = 0.48 \cdot 0.893$,
we select \texttt{park--locatedIn--city} as the pairMatch between the column pairs.
\end{example}

If a branch is selected from the existing KB, we include the granularity score in the
pairMatch score so that when we compare the score between the query table and different
data lake tables, the data lake tables matching in the granular types are prioritized
over the tables that match just on the top level types.
Our \textit{pairMatch} score may be low if there is no match at the most granular level.
One can design a different penalization scheme to penalize the frequent types to change
the trend of scoring. However, our objective is to introduce relative difference in the
match between the query table and each data lake table rather than the absolute unionability score.

Let there be $m$ matching column pairs between $Q$'s semantic tree $SG(Q)$ rooted at
intent column $I$ and $\dltable$'s subtree $SG(\dltable)$ rooted at $c$.
We compute the unionability score ($S$) between $Q$ and $\dltable$ using
\begin{multline}
\label{eq:new_unionability_score}
    S(Q,\dltable) =
    \sum_{i = 1}^{m} pairMatch((Q.I, Q.c_i), (\dltable.c, \dltable.c_i)).
\end{multline}


Intuitively, \cref{eq:new_unionability_score} favors the data lake table that has
more matching columns, finer granularity of matching types, and greater probability
of matching types.\footnote{\label{footnote:scoring_function_normalization}The score is normalized between 0 and 1 in the literature~\cite{2019_vinh_thinh_qsearch, 2021_vinh_thinh_qute}.}. The latter two are captured by the individual matching scores
while the former is addressed by the summation of such scores.
Note that unionability between two tables is often viewed as a binary decision, i.e.,
either the tables are unionable or not.
In a traditional database setting, two tables are unionable iff they have compatible
schemata.\footnote{One might also pre-process the tables for compatibility,
for example, project out columns or rename attribute names.}
In the data-lake setting, the constraint of aligning a unionable table's schema exactly
with that of a query table is not practical.
Existing works~\cite{nargesian_18_TUS, 2020_bogatu_d3l} have relaxed this in their
unionability definitions. However, they solely consider column unionability without
considering the relationships and the preservation of the query table's topics.
Thus, by introducing \SANTOS, we aim to show the importance of considering relationships
when searching for unionable tables that share the same intent as the query table.
\section{\SANTOS Implementation}
\begin{figure*}[h]
  \includegraphics[scale = 0.4]{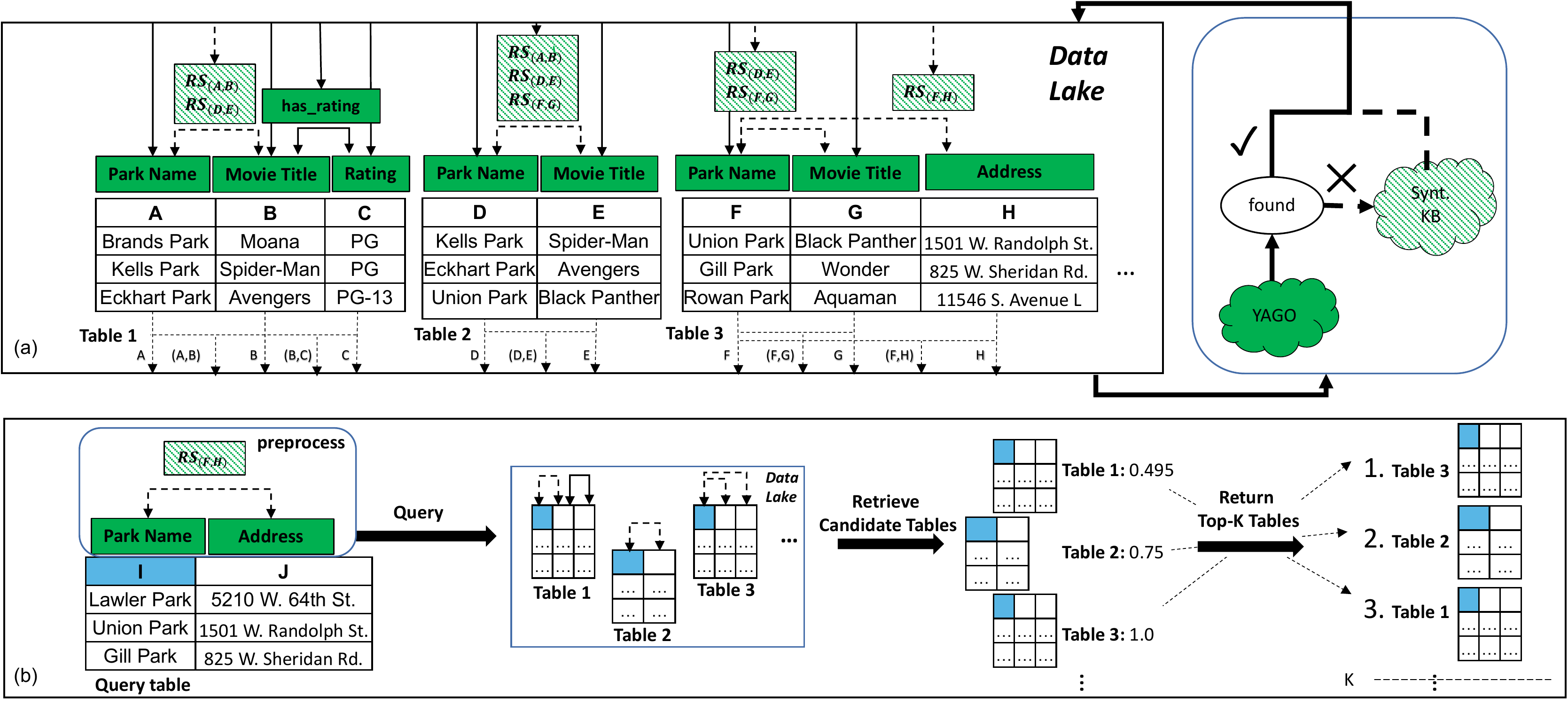}
  \caption{Pipeline of \SANTOS. (a) Preprocessing phase: data-lake tables are labeled with types from YAGO or {\em synthesized KB}. (b) Query phase: the query table is annotated, and we query the data lake to retrieve and rank unionable tables. Columns highlighted in blue represent the (matching) intent column. The type hierarchy is omitted for simplicity. 
  }
\label{fig:pipeline}
\end{figure*}
We now describe \SANTOS implementation, namely, the KB use, the synthesized KB creation, and the pre-processing and query phases.
\label{section:santos_implementation}
\subsection{Knowledge Base Implementation} 
\label{subsection:kb_details}
When creating CS and RS using YAGO types, we build four dictionaries for efficient access.
First, we store labels and alternate names (synonyms) that describe the corresponding entity, derived from YAGO, mapped to URIs\footnote{\label{footnote:labels_alternate_names}We only index labels and alternate names in English.
}, in an \emph{entity dictionary}.
\footnote{\label{footnote:rdf_labels} \url{https://www.w3.org/2004/12/q/doc/rdf-labels.html}}
\revision{
We connect tables to the KB by mapping column values to KB labels and alternate names.}
Note that YAGO permits multiple entities to have the same label or alternate name, so a value may be mapped to entities of different types (e.g. homographs~\cite{21_leventidis_homograph_detection}). 
Second, we build an 
\emph{inheritance dictionary} that stores each Top level type and their children types.
Next, we use a \emph{type dictionary} to lookup entities as keys, which are also values in the \emph{entity dictionary}, with their set of (types, granularity score) as values (e.g.
\texttt{Boston}: \{(\texttt{place}, 0.14), (\texttt{city}, 0.22),$\dots$\}).
Finally, we use a \emph{relationship dictionary} to store the set of binary relationships (properties) for each value-pair in the KB,
similar to the (proprietary) relation database used to recover semantics of webtables~\cite{venetis_11_recovering_semantics}.

As we index only the necessary KB triples, the total space taken by these dictionaries in main memory is 3.75 GB. 
We need 965.18 MB to store them persistently.
\footnote{\label{footnote:memory_size} We report main memory space returned by sys.getsizeof() 
and indexes are stored as compressed pickle files (\url{https://docs.python.org/2/library/pickle.html}).} 
\revision{
Note that, creating these dictionaries is a one-time task and takes less than 20 minutes in our experimental setup (see \cref{section:experimental_setup})}.

\subsection{Synthesized KB Implementation}\label{subsection:synth_details}
For the synthesized KB, we create a \emph{Synthesized Type Dictionary} and \emph{Synthesized Relationship Dictionary}. 
As we directly use cell values as entities, an \emph{entity dictionary} is not required. Also, the inheritance dictionary is not needed as
we consider all synthesized types to have the same granularity.
We will only discuss the creation of the \emph{Synthesized Relationship Dictionary} for column pairs, since the \emph{Synthesized Type Dictionary} is created in the same way.

\begin{figure}[ht]
  \includegraphics[scale = 0.50]{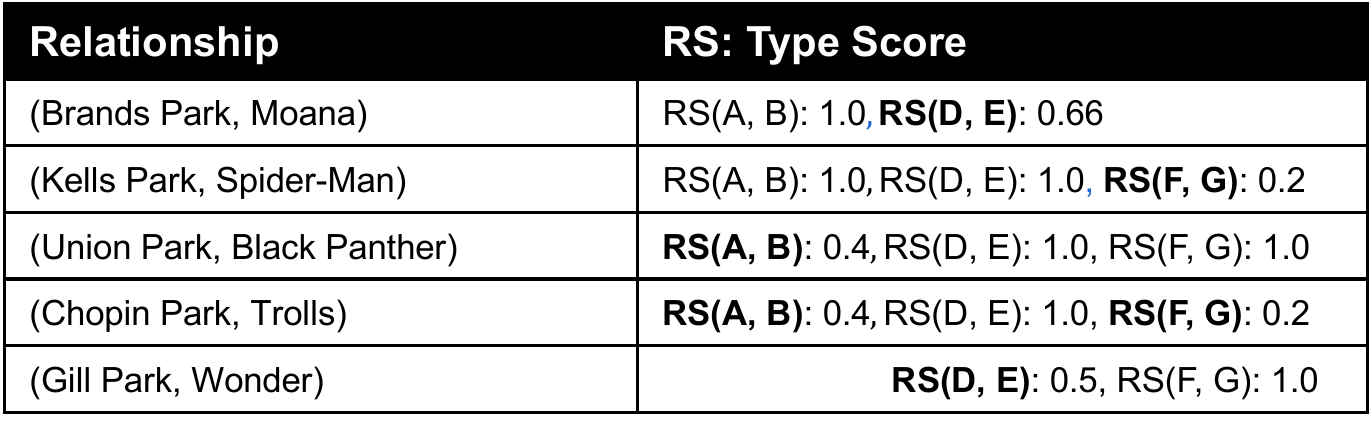}
  \caption{\emph{Synthesized Relationship Dictionary} for \cref{fig:synthetic_kb}.  
  }
\label{fig:synth_dict}
\end{figure}
\introparagraph{Synthesized Relationship Dictionary}
In \cref{section:unionability_components}, we built a \emph{relationship dictionary} using an existing KB to create annotation to add to the RS. Here, we follow an inverse process, i.e., we first assign a new binary RS to each column pair, then populate a synthesized relationship dictionary with each
\textit{value-pair} in the column pair that is not in any existing KB.
For each such value-pair, we assign a binary \emph{synthesized relationship semantic (RS)},
which is a distribution of column pairs that exhibit some meaningful relationships among value-pairs. 
A similar idea exists in topic modeling~\cite{papadimitriou2000latent}, where some latent topic of a document (value-pair) is represented by a distribution over other documents (column-pairs).

We first assign all value pairs in a column pair with the same synthesized RS. 
For instance, value-pairs \texttt{(Brands Park, Moana)}, \texttt{(Kells Park, Spider-Man)} and \texttt{(Eckhart Park, Avengers)} in Table 1 of \cref{fig:synthetic_kb} are all assigned $RS{(\texttt{A,B})}$. 
As we have discussed however (\cref{ex:relationshipsemantics}), some binary relationships between columns in a data lake are not meaningful and may be indirect. For example, in \cref{fig:running_examples} Table (c), value-pairs in $\Pi_{Film Title, Park Location}$ 
like \texttt{(Bee Movie, 6748 N. Sacramento Ave.)} and
\texttt{(Coco, 5210 W. $64^{\mathrm{th}}$ St.)},
have a tenuous relationship at best (film shown in a park with this address). Thus, we aim to capture only \emph{meaningful} binary relationships in which one column functionally determines another.
For example, in \cref{fig:running_examples}(b), \texttt{Park Name} functionally determines \texttt{Park Location}, indicating a possible meaningful relationship. 
We hypothesize that column pairs in a functional dependency are more likely to contain a meaningful semantic relationship that also exists in other tables and may be useful in union search. 
This also has a benefit of reducing the synthesized relationship dictionary size. 
We use an existing 
\emph{functional dependency} (FD) discovery algorithm called FDEP~\cite{1999_flach_database_dependency_discovery} to find unary FDs (FDs with a single column determinant in binary relationships),
and run their bottom-up variant that first considers all pairwise relationships, then checks if a dependency satisfies an FD. 
Although FDEP time complexity is quadratic,
we only use it offline.

To generate our synthesized relationship dictionary, we first iterate over the data lake tables and store value-pairs in column pairs (not found in the existing KB) that form an FD, and their synthesized RS with a type score of 1, in a \emph{lookup dictionary}. We then find the type score of each RS to capture the likelihood for the associated value-pair to have that semantics. Type scores, also seen as confidence scores for value-pairs, are also calculated based on the overlap of value pairs in different tables (consistent with \cref{eq:relation_synth_score}). However, we now consider value-pairs rather than column-pairs, so we calculate the overlap score for each value-pair in a column pair. Thus, we fill the \emph{Synthesized Relationship Dictionary (S)} with RS from the lookup dictionary and their associated type scores. We clarify the computation using the following example.



\begin{example}
\label{example:synth_kb_scores}
Consider Tables 1, 2, and 3 in \cref{fig:synthetic_kb} and S in \cref{fig:synth_dict}. 
First, we assign all value-pairs with 
$RS{(\texttt{A,B})}, RS{(\texttt{D,E})}$ and $RS{(\texttt{F,G})}$ with respect to their consisting column pairs with type score 1. 
Next, consider Table 1
($RS{(\texttt{A,B})}$), which shares two value-pairs (bolded) with Table 2 ($RS{(\texttt{D,E})}$) out of three total value-pairs. 
\texttt{(Brands Park, Moana)} in Table 1 is then also assigned $RS{(\texttt{D,E})}$ with a type score of $\frac{2}{3}$.
Now consider Table 2, which shares two value-pairs with Table 1 (bolded)
and one value-pair with Table 3 (italicized),
out of five total value-pairs. Then, \texttt{(Chopin Park, Trolls)} is assigned $RS{(\texttt{A,B})}$ with a type score of $\frac{2}{5}$ = 0.4 and $RS{(\texttt{F,G})}$ with a type score of $\frac{1}{5}$ = 0.2. 
\end{example}

\subsection{Pre-Processing Phase}
\label{section:preprocessing_phase}
During the offline pre-preprocessing phase, 
we find CS and RS for data lake tables using an external KB (\cref{section:unionability_components}),
and discover FDs in the data lake tables to create a synthesized KB (\cref{subsection:synth_details}).
To reduce query time, we create two inverted indexes.
First, a \textit{node inverted index} maps a column to its CS with respective $\CSconf$. 
The second is an \textit{edge inverted index} that maps RS to its connected columns in the same table, with $\RSconf$.

\introparagraph{Time Complexity}
Consider a set of data lake tables $\mathcal{\dltable}$ and let $m$ and $n$ be the largest number of columns and rows respectively in a data lake table.
We make a linear pass over each column and record the count of each candidate CS (include candidate top-level types) in an inverted index. 
So, the asymptotic time complexity of computing CS is $\O(|\mathcal{\dltable}| \cdot m \cdot n)$. 
Similarly, for computing RS, we only consider columns with non-empty CS, which can then find RS in the KB~\cite{tanon_20_yago}. 
Let $m_c$ be the largest number of columns having CS in a data lake table such that $m_c \leq m$. The time taken to compute RS is bounded by  $\O(|\mathcal{\dltable}| \cdot m_c^2 \cdot n)$.

The time complexity of creating synthesized KB includes the creation of \emph{Synthesized Type Dictionary} and \emph{Synthesized Relationship Dictionary}.
\emph{Synthesized Type Dictionary} creation is analogous to CS computation, where we make a linear pass over each column ($\O(|\mathcal{\dltable}| \cdot m \cdot n)$ time). 
Also, the creation of \emph{Synthesized Relationship Dictionary} is analogous to RS computation. 
But in addition, we also need to mine unary FDs, which is quadratic in the number of columns~\cite{1999_flach_database_dependency_discovery}. 
So, its creation is linear in the highest number of rows and quadratic only in the highest number of columns across any table i.e., $\O(|\mathcal{\dltable}| \cdot m^2 \cdot n)$. 
Recall that this is a pre-processing task and does not need to be done during the query phase. We report the pre-processing time on different benchmarks in \cref{section:results_effeciency}.

\subsection{Query Phase}
\label{section:query_phase}
At query time, shown in \cref{fig:pipeline}(b), the user provides a query table $Q$ in which the intent column
is marked.\footnote{The user may also select a set of property columns to consider. In our experiments (\cref{section:experiments}), we simply assume that the user is interested in all property columns.}
We then create the semantic tree for $Q$,
rooted at the intent column. 
For this, we access KB and synthesized KB to find the CS and RS for the query table.
Given the annotated semantics of the query table, \SANTOS searches for unionable tables.
It does so by 
retrieving a set of candidate tables and their respective confidence scores from the inverted index. 
\SANTOS  computes the unionability score between column pairs of $Q$ and candidate tables based on different hierarchy levels, 
allowing us to match columns of different granularities (see \cref{eq:col_match}).

\introparagraph{Time complexity} 
Given a query table $Q$ with m columns and n rows, we compute its CS from both KB and synthesized KB by making a single pass over each column, which takes $\O(m \cdot n)$ time. Recall that we only find RS for columns that have non-empty CS. Let $m_c$ be the number of columns having CS such that $m_c \leq m$. RS can thus be computed in $O(m_c^2 . n)$ time. After finding CS and RS for the query table, the computation of unionability score depends linearly on the number of semantics found for the query table. With inverted indexes and the number of possible CS and RS as constant during the query phase, it is efficient to compute the unionability score. We report the query time using real data lake tables in \cref{section:results_effeciency}.

\section{Experiments}
\label{section:experiments}


In this section, 
we compare \SANTOS with 
a
state-of-the-art unionability approach $D^3L$~\cite{2020_bogatu_d3l} that uses column unionability.
Notice that $D^3L$ builds on Table Union Search~\cite{nargesian_18_TUS} by adding regular expressions, domain distributions, etc.  Hence, we only compare with $D^3L$ to analyze the importance of relationship semantics in union search. 
\subsection{Experimental Setup}
\label{section:experimental_setup}
\SANTOS is implemented in Python on a server with
Intel(R) Xeon(R) Gold 5218 CPU @ 2.30GHz
processor. Our code is publicly available.\footnote{\label{footnote:santos_github}\url{https://github.com/northeastern-datalab/santos}}
%
Our experiments 
 aim to answer the following questions: 


\begin{enumerate}
    \item How effective is \SANTOS in returning \topk unionable tables relative to the baseline given a query table? (\cref{section:results_effectiveness})

    \item How do each component of \SANTOS (use of existing KB, use of synthesized KB, and use of both) influence 
    the quality of the results? (\cref{section:results_KB})

    \item How well does \SANTOS scale
	over real data lakes, as compared to prior work? (\cref{section:results_effeciency})

\end{enumerate}

\introparagraph{Evaluation Measures}
Since our method, along with other table union search methods, returns a \topk list of unionable tables, we use mean average precision ($MAP@k$) to evaluate the effectiveness of table union search approaches~\cite{manning2008introduction}.\
Following previous works~\cite{nargesian_18_TUS,2020_bogatu_d3l}, we also report Precision at $k$ ($P@k$) and Recall at $k$ ($R@k$)~\cite{manning2008introduction}.
When creating our ground truth for evaluations, we assign a binary score $\in \{0, 1\}$ to a data lake table $T$, given a query table $Q$ to label $T$ as unionable or not-unionable to $Q$. Formally, let $\mathcal{T}_{Q}$ be the set of unionable tables based on the ground truth and $\hat{\mathcal{T}}_{Q}$ be the set of \topk unionable tables based on some method with respect to a query table $Q$. Then, $P@k$ and $R@k$ with respect to $Q$ are given by:

\begin{equation}\label{eq:P_R}
    P@k = \frac{\mathcal{T}_{Q}\cap\hat{\mathcal{T}}_{Q}}{\hat{\mathcal{T}}_{Q}}, R@k = \frac{\mathcal{T}_{Q}\cap\hat{\mathcal{T}}_{Q}}{\mathcal{T}_{Q}}
\end{equation}

Note that the size of $\hat{\mathcal{T}}_{Q}$ is set to $k$, while $\mathcal{T}_{Q}$ may, in general, be larger (or smaller) than $k$. To best understand the results, we create benchmarks where the ground truth ($\mathcal{T}_{Q}$) is at least $k$. Hence, using $P@k$, if a method returns less than $k$ results, the results not returned are considered incorrect (false). For instance, if $k=10$ and the ground truth has 20 results, if a method returns only 2 results out of which 1 is correct and the other is incorrect, then $P@10 = \frac{1}{10}$ and $R@10 = \frac{1}{20}$. So when $k < |\mathcal{T}_{Q}|$ then perfect recall is not possible as the $R@k$ can at best be $\frac{k}{|\mathcal{T}_{Q}|}$.

The mean average precision ($MAP@k$), defined as follows:
\begin{equation}\label{eq:map}
    MAP@k = \frac{1}{|\hat{\mathcal{T}}_{Q}|}\sum_{k=1}^{|\hat{\mathcal{T}}_{Q}|} P@k
\end{equation}



We compute $P@k$, $R@k$, and $MAP@k$ for each query and report the average performance over several queries for a fixed $k$ (e.g. average $P@k$).
We also measure the
pre-processing and query times for scalability experiments. 

\begin{figure}[h]
    \centering
  \includegraphics[width=\linewidth]{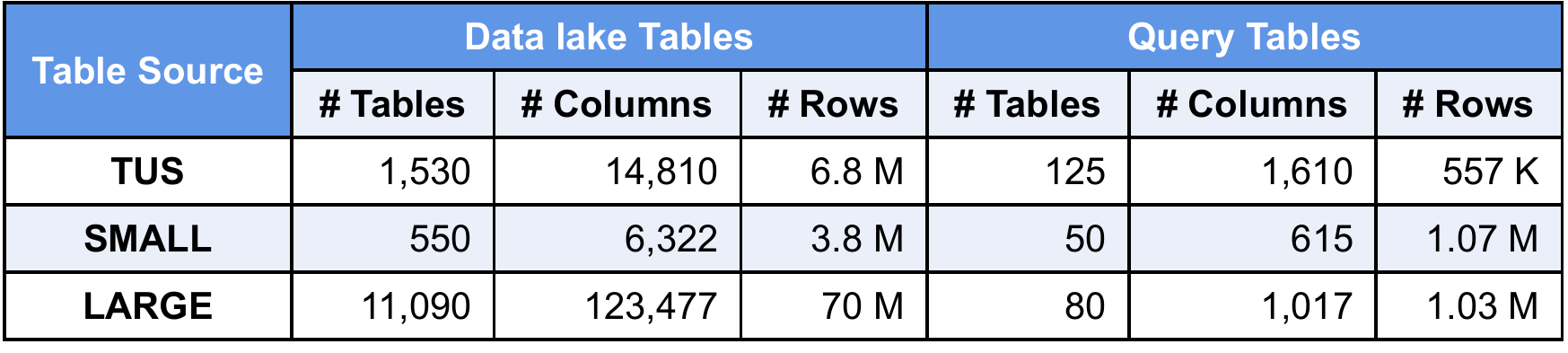}
  \caption{Benchmarks used in the experiments.}
\label{fig:benchmark_statistics}
\end{figure}

\introparagraph{Benchmarks}
\Cref{fig:benchmark_statistics} details the statistics of the benchmarks that we use,
both for their data lake tables and the query tables. The benchmarks are publicly available.
\footnote{\label{footnote:santos_github}\url{https://github.com/northeastern-datalab/santos/tree/main/benchmark}}
\emph{{1. TUS Benchmark (TUS)}}: 
The TUS benchmark~\cite{nargesian_18_TUS}
focuses on attribute unionability, ignoring relationships between columns.
Specifically, two tables in the benchmark are unionable if they have unionable attributes. 
To repurpose this benchmark, we labeled tables in the benchmark unionable if they share relationships, not just attributes.
Specifically, out of the 10 original seed tables used to produce 1530 tables in the benchmark, 
we found 
meaningful relationships in tables that originate from 6 seed tables. 
From these, we randomly selected 125 tables
as Query Tables
and marked the intent columns. 
We use the original data lake from the benchmark. 



\emph{{2. SANTOS Small Benchmark (SMALL)}}:
The SANTOS Small Benchmark
contains 550 data lake tables from Canada, UK, US, and Australian open data.
First, we collected 296 real data lake tables from 35 distinct domains. 
To further expand the benchmark, we selected 19 large tables among them and manually annotated the relationships between their columns. 
Using the benchmark creation technique from TUS~\cite{nargesian_18_TUS}, we then partitioned the annotated tables horizontally and vertically to obtain 254 non-overlapping tables, thereby increasing the total number of tables to 550. 
Then we randomly selected 50 tables having at least 10 unionable tables as query tables (at most 2 query tables from each domain) and labeled the intent columns and the ground truth. 
As the tables are taken from real data lakes, they contain nulls, string values, date, numerical values, etc. We will refer to this benchmark as \emph{SMALL}.

\emph{{3. SANTOS Large Benchmark (LARGE)}}:
To evaluate SANTOS in a broader environment, we collected 11,090 real tables from Canada and UK Open Data for the data lake.
From these tables, we randomly selected 80 tables as query tables, each having at least 20 unionable tables, and marked their intent columns.
For this benchmark, we only report $P@k$ and $MAP@k$ (and runtimes) as reporting $R@k$ requires a laborious annotation of the full data lake. For similar reasoning, we manually verify $P@k$ and $MAP@k$ only up to $k=20$.
Like SMALL Benchmark, these tables also contain nulls, string values, dates, numerical values, etc.
We refer to this benchmark as \emph{LARGE} from now on.
\introparagraph{Baselines}\label{section:baselines}
The table union search problem based on determining if column values are drawn from the same domain was first defined and addressed by Nargesian et al.~\cite{nargesian_18_TUS}. 
Recently, Bogatu et al. proposed $D^3L$ 
for the broader problem of finding related tables (both joinable and unionable tables)~\cite{2020_bogatu_d3l}. 
$D^3L$ adds metrics based on column names, regular expressions and domain distributions to the word-embedding and value overlap-based models
of
Nargesian et al.~\cite{nargesian_18_TUS}.
Therefore, we compare \SANTOS to $D^3L$ with these extended metrics,
by reproducing $D^3L$ using their code.\footnote{\url{https://github.com/alex-bogatu/d3l}}

TURL is a 
recent method that uses representational learning over web tables~\cite{2020_deng_turl}. 
TURL learns table
representations that successfully find CS (column type annotation) and RS (relation extraction) in {web tables}~\cite{2020_deng_turl}.
Although it does not support union search directly, we extended it to create a \SANTOS-like technique. 
Specifically, we treat TURL as a KB and, 
similar to \SANTOS, 
use it to annotate the CS and RS
for each table. Then we index the data lake tables similarly
to the method in \cref{section:unionability_components}. 
This approach provides an analysis of a learning-based alternative that uses a pre-trained model.

\subsection{\SANTOS Effectiveness vs. Baselines}\label{section:results_effectiveness}
To analyze the effectiveness of \SANTOS, we compare $P@k$, $R@k$
and $MAP@k$ 
of \SANTOS against our baselines $D^3L$ and TURL.
We then analyze variations of \SANTOS
in an ablation study in \cref{section:results_KB}.



\begin{figure}[h]
    \centering
  \includegraphics[width=0.75\linewidth]{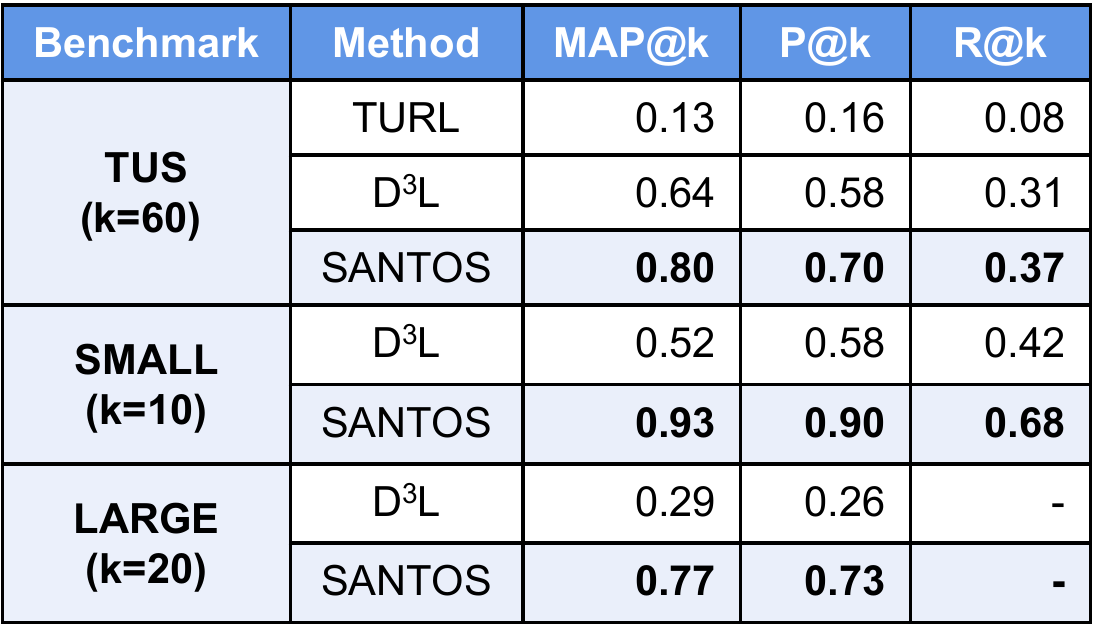}
  \caption{Comparison of  
  P@k, MAP@k and R@k of TURL, $D^3L$ and \sFull on TUS, and $D^3L$ and \sFull on SMALL and LARGE benchmarks.
  }
\label{fig:pr_results}
\end{figure}

\Cref{fig:pr_results} 
reports the average performance of \SANTOS compared to the baseline $D^3L$ on all three benchmarks.
We only report TURL for TUS benchmark, since its performance is similar on the other benchmarks, with its measures significantly less than \SANTOS.
For consistency with previous
research~\cite{nargesian_18_TUS,2020_bogatu_d3l}, and to ensure that each query table has at least $k$ expected unionable tables, 
we report results for $k=60$ on TUS. Considering the number of unionable tables per query table, we report results at $k = 10$ on SMALL and at $k = 20$ on LARGE (\cref{fig:pr_results}).
Note that when the ground truth contains more than $k$ results, 100\% $R@k$ is not possible.  Rather the highest possible $R@60$ is around 62\% for TUS and $R@10$ is around 72\% for SMALL (Ideal lines in (b) and (d)).  In the LARGE data lake the whole corpus is not labeled. Therefore, we cannot report $R@k$.

\begin{figure}[h]
    {
    \centering
    \begin{minipage}[t]{\textwidth}
    \includegraphics[width=0.48\linewidth]{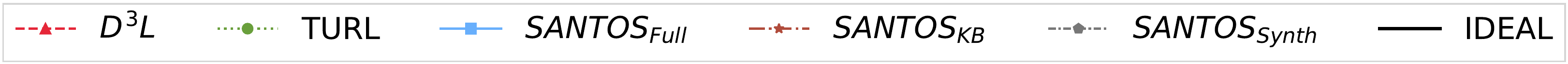}
    \end{minipage}%
    }
    \subfloat[Average $P@k$ on TUS]{
    \begin{minipage}[t]{0.49\linewidth}
    \includegraphics[width=\linewidth]{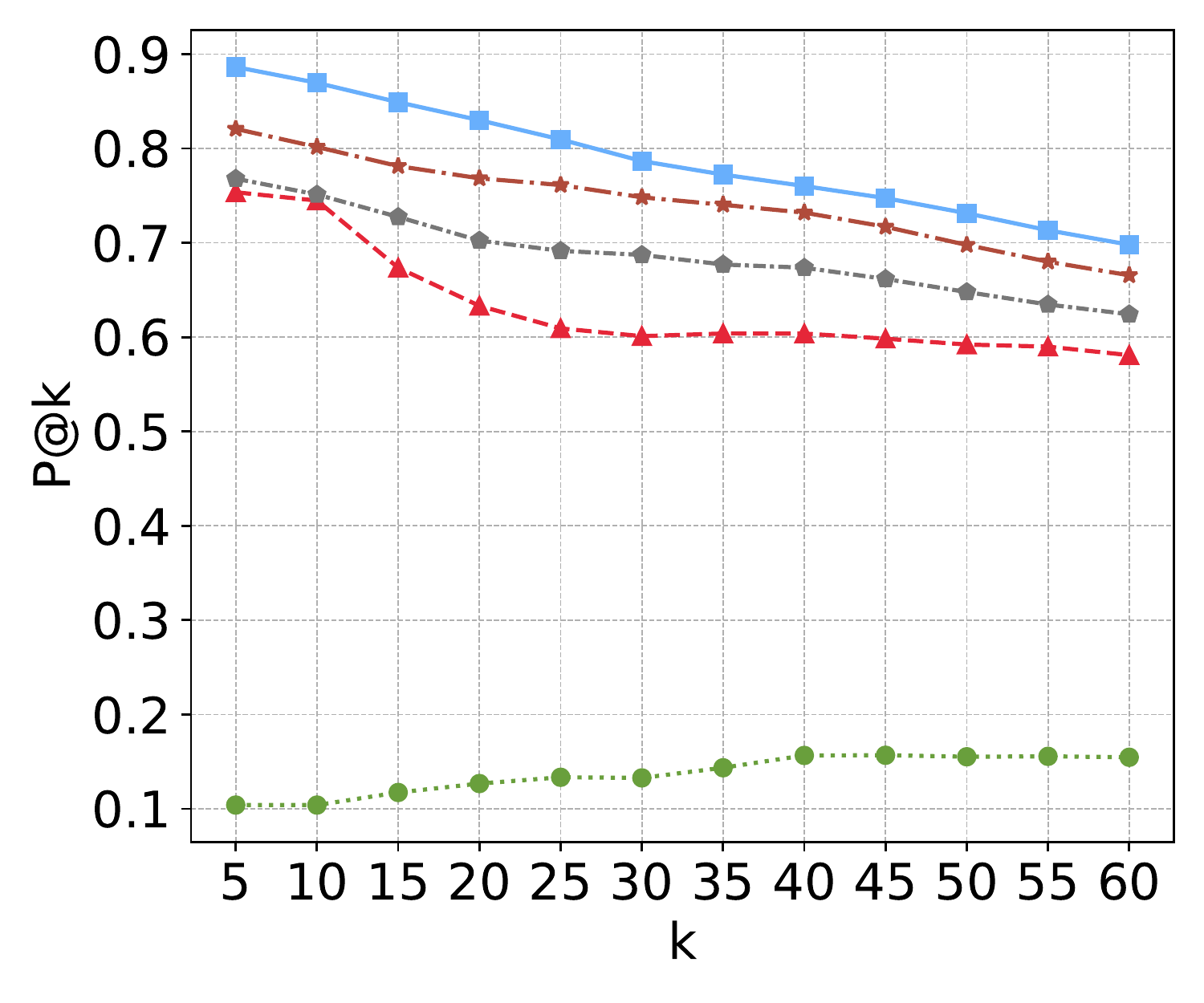}
    \end{minipage}%
    }
    \subfloat[Average $R@k$ on TUS]{
    \begin{minipage}[t]{0.49\linewidth}
    \includegraphics[width=\linewidth]{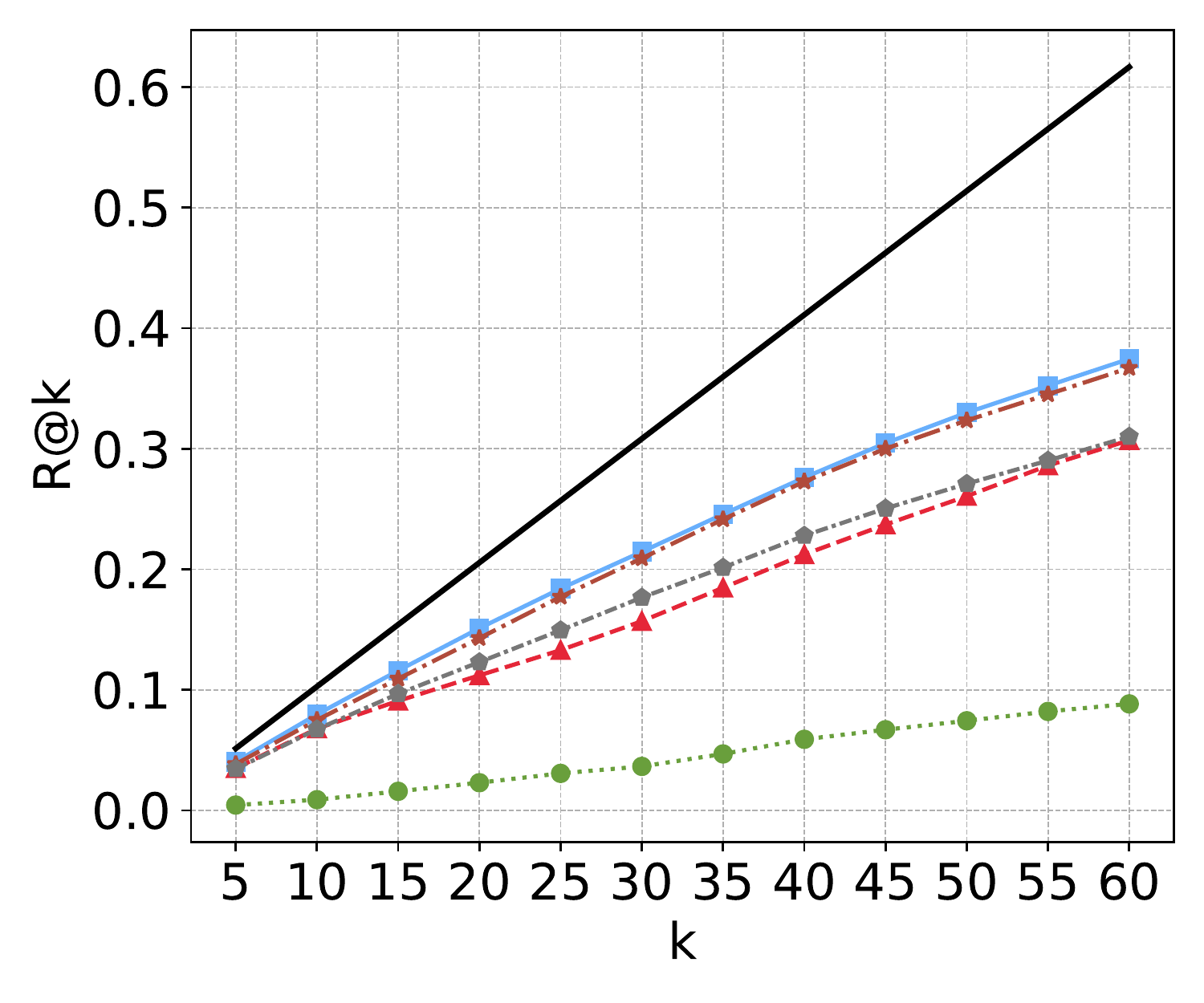}
    \end{minipage}%
    }
    \hfill
    \subfloat[Average $P@k$ on SMALL]{
   \begin{minipage}[t]{0.49\linewidth}
    \includegraphics[width=\linewidth]{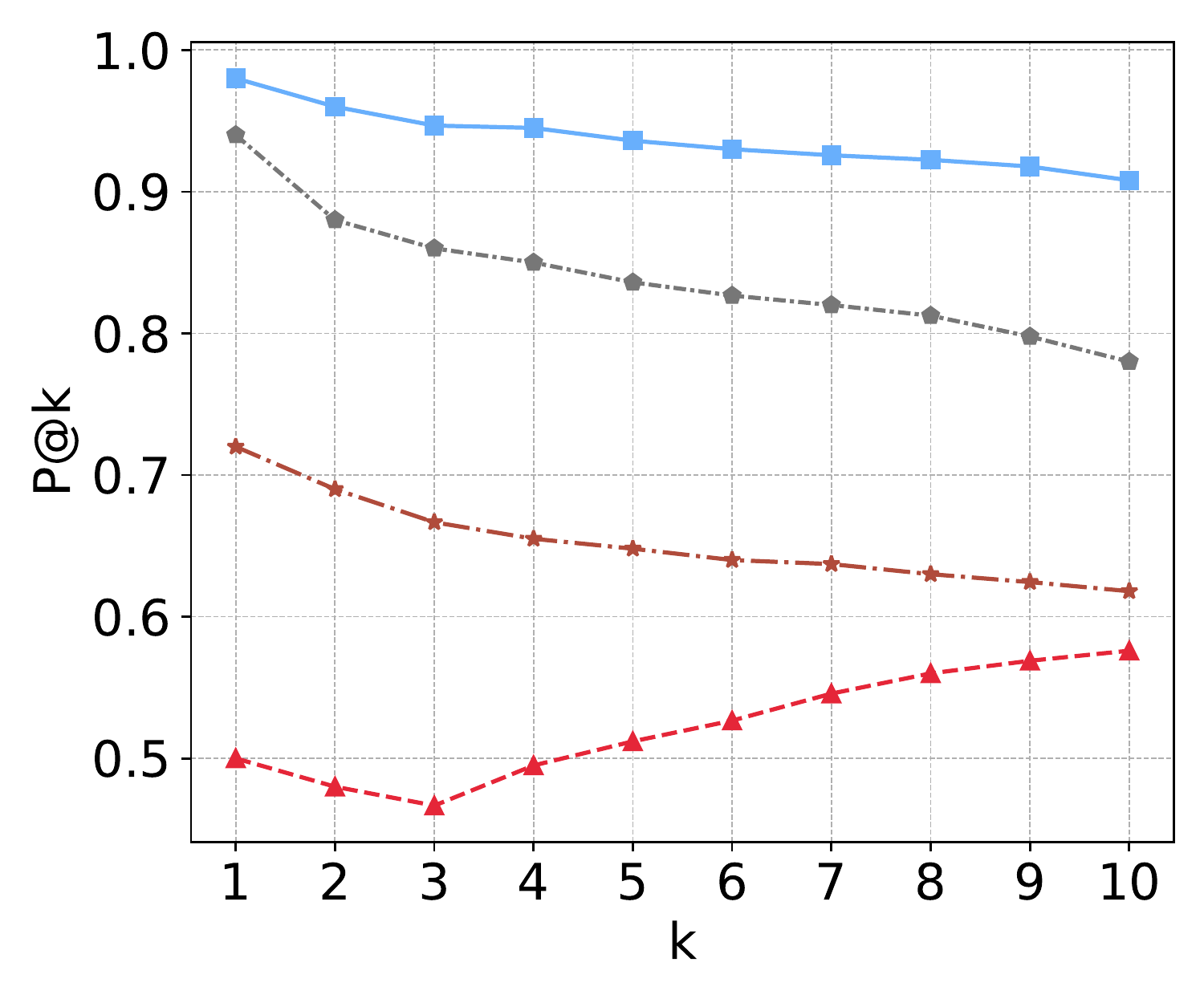}
    \end{minipage}%
    }
    \subfloat[Average $R@k$ on SMALL]{
    \begin{minipage}[t]{0.49\linewidth}
    \includegraphics[width=\linewidth]{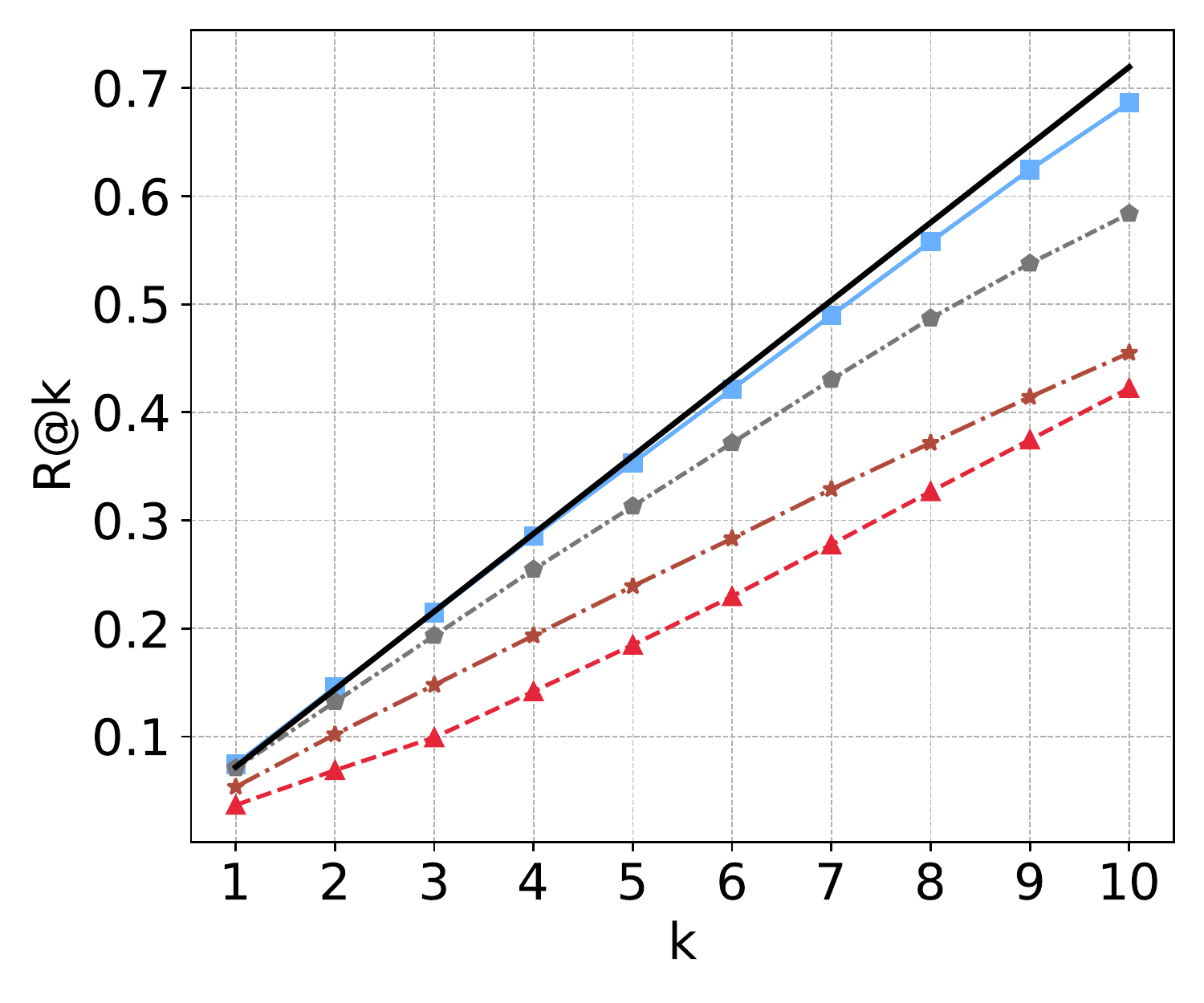}
    \end{minipage}%
    }
    \caption{Effectiveness of \SANTOS and its variations against baselines in different benchmarks
	}
    \label{fig:pr_rc_all}
\end{figure}

We first evaluated a version of SANTOS that only considers column semantics, \sCol, compared against $D^3L$. On TUS, \sCol has $MAP@60$ of 65\% and $P@60$ of 62\%, which is comparable to $D^3L$ results. When we included relationship semantics and evaluated full \SANTOS against $D^3L$, however, \SANTOS performs better than $D^3L$ over all measures in all benchmarks. 
On TUS, \SANTOS outperforms $D^3L$ by over 20\%, 25\% and 19\% in terms of $P@60$, $MAP@60$, and $R@60$, respectively.
On SMALL, \SANTOS outperforms $D^3L$ by over 56\%, 78\%, and 61\% for $P@10$, $MAP@10$, and $R@10$ respectively. 
In terms of $R@k$, \SANTOS is closer to ideal recall than the baseline in both benchmarks.
This indicates that relationship semantics is important in union search.

On a real data lake (LARGE), we observe even further improvements ($P@20$ and $MAP@20$ by over 180\% and 165\% compared to $D^3L$
respectively), indicating that relationships are even more important in this benchmark.
For illustration, consider a query table $Q$ in LARGE benchmark about biodiversity in different counties (with columns like \texttt{county\_name}, \texttt{animal\_scientific\_name}, \texttt{documented\_year}, etc.). 
For this table, the \topk results by \SANTOS contain
tables about alpine birds, fish, and trees and the places they are found, which seems to be correct. 
However, as  discussed in \cref{example:intent_column_importance}, although tables returned by $D^3L$ have common (unionable) columns, they are about different topics with different relationships.
For instance, they include tables about emergency hospital admissions after accidents
because they contain 
columns like \texttt{county} (unionable with \texttt{county\_name} in $Q$) that describes the place of accident and \texttt{year} (unionable with \texttt{documented\_year}) that describes when the accident took place. 
For both approaches,
we manually verify $P@k$ and $MAP@k$ up to $k=20$.
The raw results for each query table on LARGE by SANTOS and $D^3L$
are available in the supplementary materials.
$^{\ref{footnote:santos_github}}$

\Cref{fig:pr_rc_all} shows detailed comparison of precision and recall for different values of $k$ in both TUS and SMALL benchmarks. 
In these graphs, \SANTOS is labeled \sFull.  The other lines labeled \SANTOS will be explained below. For each $k$, \SANTOS outperforms competing methods.  Notice that perfect $R@k$ is also plotted in \cref{fig:pr_rc_all}(b) (solid black line). TURL-based implementation has the least precision ($P@60 = 0.15$), MAP ($MAP@60 = 0.1$) and Recall ($R@60 = 0.1$) on TUS. 
It is possible that the reason TURL performs poorly is that is trained over web tables, which have different characteristics from real open data~\cite{DBLP:journals/pvldb/Miller18}. 
We only report TURL performance on TUS as the results in other benchmarks show a similar trend that is
 well below $D^3L$ and \SANTOS.

\subsection{\SANTOS Effectiveness Ablation Study}\label{section:results_KB}
\begin{figure}[t]
    \centering
    \subfloat[MAP@60 on TUS]{
    \begin{minipage}[t]{0.45\linewidth}
    \includegraphics[width=\linewidth]{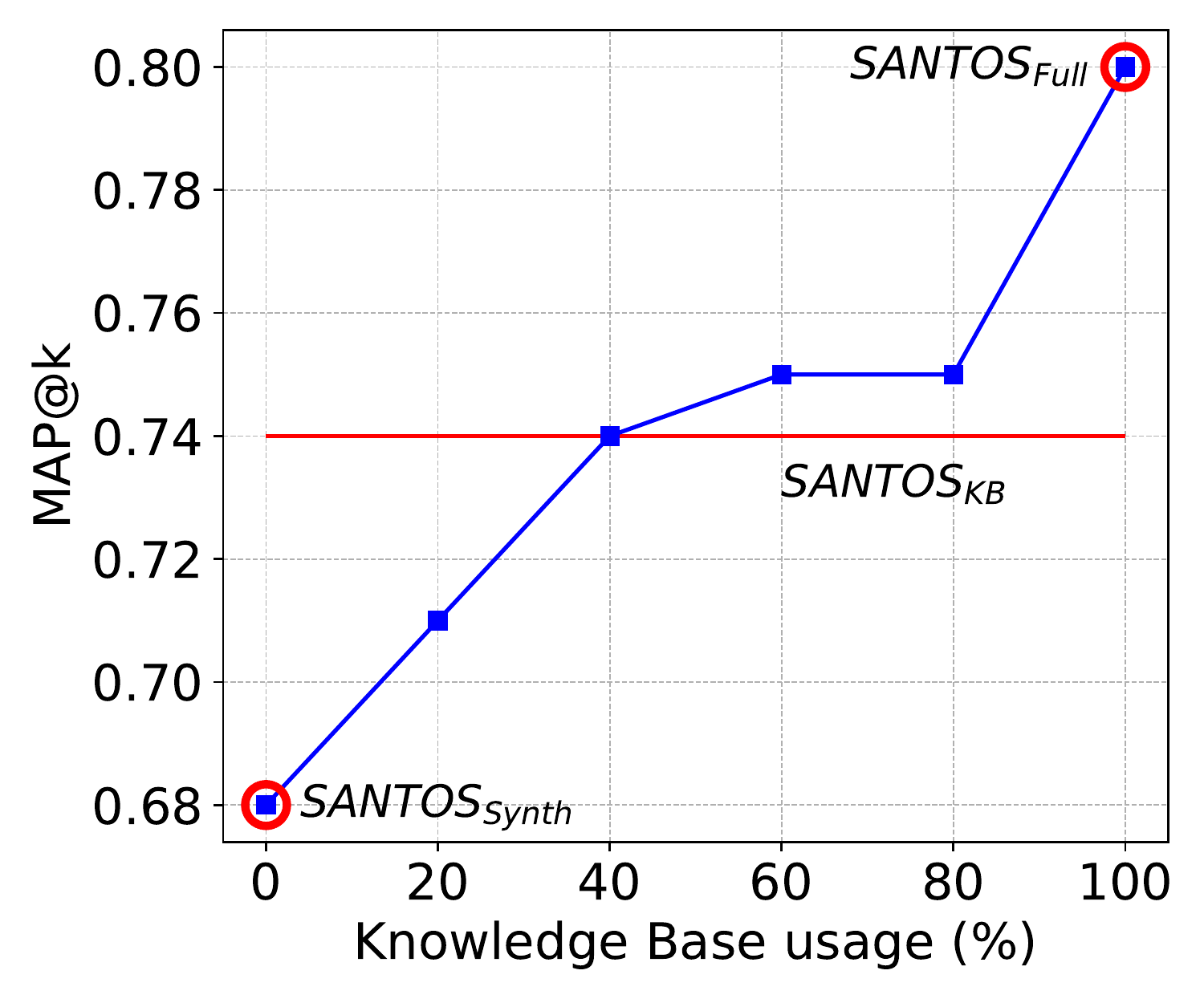}
    \end{minipage}%
    }
    \centering
    \subfloat[MAP@10 on SMALL]{
    \begin{minipage}[t]{0.45\linewidth}
    \includegraphics[width=\linewidth]{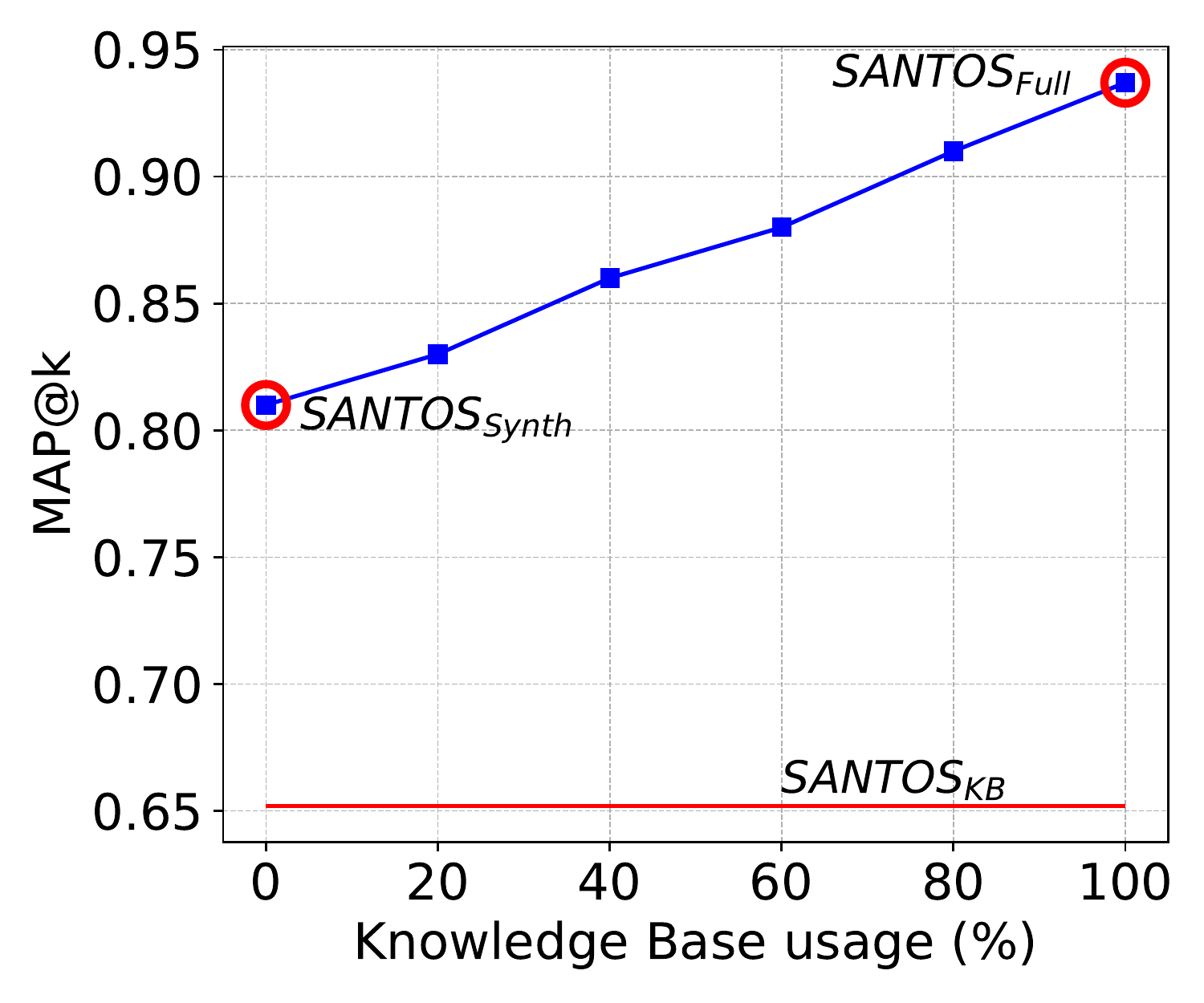}
    \end{minipage}%
    }
    \caption{Average $MAP@k$ of \sFull ($k=60$ on TUS, $k=10$ on SMALL) for different percentages of the existing KB}
    \label{fig:variable_kb_usage}
\end{figure}


We perform an ablation study to understand the impact of one of our key innovations, the use of a synthesized KB, on the accuracy of \SANTOS.
We compare the full version of \SANTOS (\sFull), a version of using solely the existing KB (\sKB), and a version only using the synthesized KB (\sSynth) over varying values of $k$ on TUS and SMALL benchmarks. 
\cref{fig:pr_rc_all} shows these comparisons in 
detail.
\sFull obtains the best results over all values of $k$ in both benchmarks.
On TUS (\cref{fig:pr_rc_all} (a) and (b)), 
\sSynth (circle/grey line) has lower $P@k$ and $R@k$ than \sKB because it is not able to retrieve enough results for all the query tables. 
However, using the synthesized KB with the existing KB together (\sFull) provides the best performance.
We see a different trend on SMALL (\cref{fig:pr_rc_all} (c) and (d)) where \sKB  was not able to return enough results. Specifically, the existing KB had no coverage for 14 of the 50 query tables. However, \sSynth was able to
handle those queries and hence, maintain the overall performance of \sFull.  
This suggests that the synthesized KB helps
\SANTOS alleviate the effect 
from the imperfect coverage of a KB. Note that we ran our experiments using an open KB over  open data tables. YAGO may cover less (or more) entities in an enterprise data lake, but 
enterprises generally maintain their own domain-specific KB's~\cite{2021_yahya_knowledge_graphs_4, 21_hogan_knoweledge_graph_survery, 2021_zalmout_product_knowledge_graph}.  \SANTOS can easily be adapted to such data lakes by augmenting YAGO with the respective enterprise KBs. 
To better understand the contribution of the KB and synthesized KB, we compute the average $MAP@k$ of \sFull by varying the percentage of information from the existing KB randomly. 
We 
systematically remove portions of the existing KB entities that are in the data lake tables
and evaluate how the synthesized KB compensates for the loss of KB coverage.
\cref{fig:variable_kb_usage} shows this analysis on the TUS  and SMALL benchmarks for $k=60$ and $k=10$, respectively. We first turn off the existing KB and compute $MAP@k$ for \sSynth. Then we gradually increase the existing KB coverage until we reach 100\%.
In both benchmarks, 
increasing the usage of the  KB increases \SANTOS's effectiveness 
almost linearly, empirically validating that \SANTOS performs better with more KB coverage.
Furthermore, it shows the significance of both \sKB and \sSynth where, \sKB increases \sFull's $MAP$ by 18\% on TUS and 15\% on SMALL and \sSynth increases it by 8\% on TUS and 43\% on SMALL.
This shows that \sSynth alone has decent accuracy (68\% on TUS and 81\% on SMALL), and incrementally adding entries from the KB improves the accuracy further in a near-linear trend. Thus, leveraging relationship semantics for union search is benefited by the use of both KB's.

Recall that, \SANTOS only uses cell values to create the semantic graphs. So, the accuracy of the created graph may depend on the number of rows in the query tables i.e., fewer rows may impact its accuracy. However, even a human expert would need an adequate number of rows to understand the table semantics and so does \SANTOS.
As reported in \cref{fig:benchmark_statistics}, the data lake tables generally contain thousands of rows and hence, \SANTOS is fairly effective in understanding their semantics as reported in our experiments.



\subsection{\SANTOS Scalability}\label{section:results_effeciency}
\begin{figure}[h]
    \centering
  \includegraphics[width=0.85\linewidth]{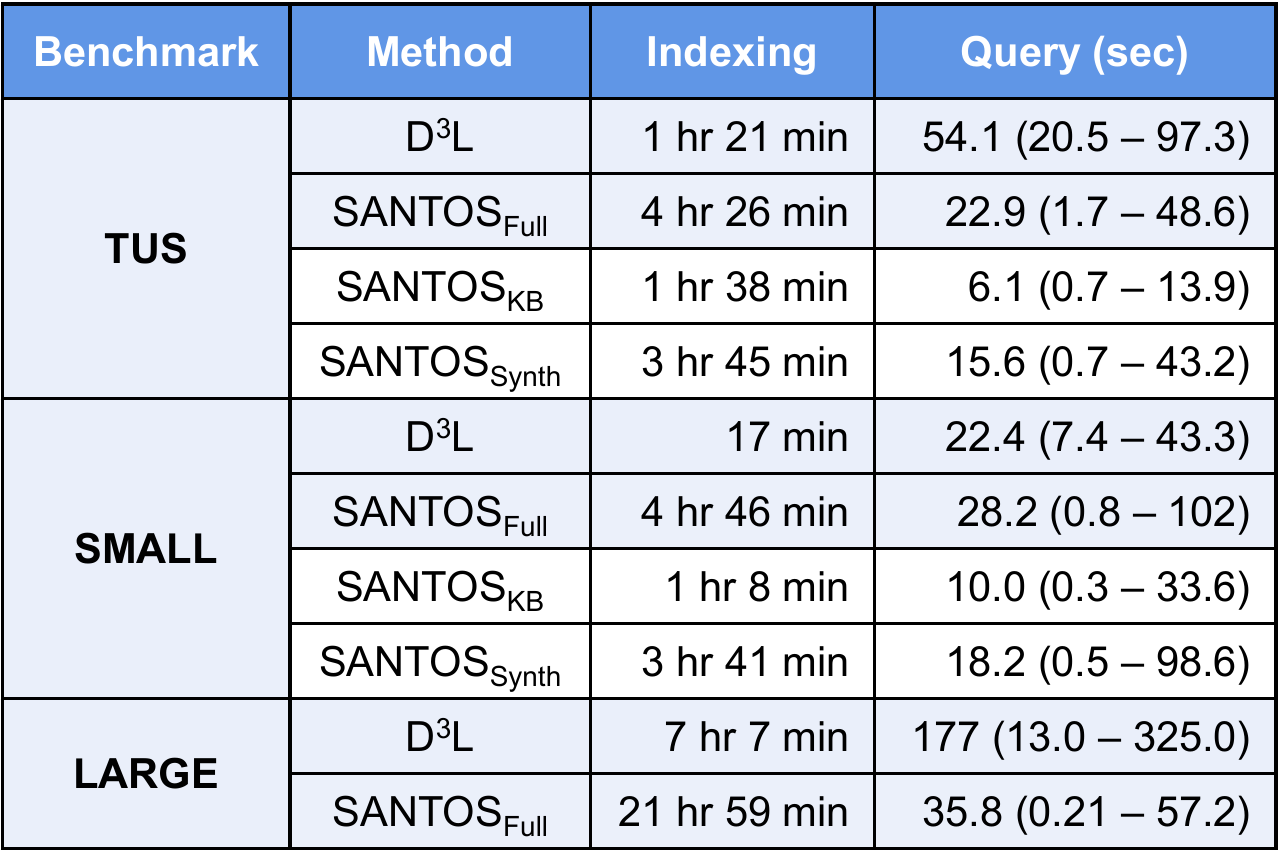}
  \caption{Comparison of Indexing and Query times of \sSynth, \sKB, \sFull, and $D^3L$ in different benchmarks.
  For Query time, we report both the average query time and in parenthesis we include 80\% confidence interval (10\% to 90\% percentile).
 }
\label{fig:efficiency_time}
\end{figure}


As a final analysis, we experimentally validate that \SANTOS scales to large data lakes. We report indexing and query times for \SANTOS (\sFull), 
the two variations
of \SANTOS (\sKB and \sSynth) and $D^3L$ on all three benchmarks (\cref{fig:efficiency_time}). For discussion, we focus on LARGE as it is largest in size.

Although $D^3L$ is around 3x faster than \SANTOS when indexing data lake tables (\cref{section:preprocessing_phase}), the relationship-based approach of \SANTOS proves to be significantly more effective than the column-based approach of $D^3L$.
Nevertheless, \SANTOS is much faster than $D^3L$ at query time (\cref{section:query_phase}).
In LARGE, \SANTOS takes 7 sec on average to index each data lake table, and thus is able to
scale to tables with thousands of rows and columns (see \cref{fig:benchmark_statistics}).
The individual index creation time of \sSynth, which includes the time taken to discover FDs, is 17 hr 19 min. 
Recall that \SANTOS may be further effective if the tables do not contain homographs. Using the state-of-the-art technique~\cite{21_leventidis_homograph_detection}, the time taken to detect the homographs is 
17 min on the TUS Benchmark. Therefore, we can detect homographs in fairly small and feasible pre-processing time. Notice however, the absence of homographs is not a necessary condition and \SANTOS is still more effective than baselines without this pre-processing step (see \cref{section:results_effectiveness}).
Similarly, \sKB's index creation time is 12 hr 40 min. 
One can create these indexes in parallel, so \sFull indexing time can alternatively be max(17 hr 19 min, 12 hr 40 min) = 17 hr 19 min plus few seconds to combine the indexes (rather than 21 hr 59 min reported in \cref{fig:efficiency_time}).
The synthesized KB created over LARGE uses 3588 MB of main memory at query time and occupies 3441 MB in the secondary storage. $^{\ref{footnote:memory_size}}$ 
This further validates \SANTOS's scalability. In the future work, we will study other spaces for optimizing the creation of synthesized KB.

We analyze 125 Query Tables on TUS, 50 Query Tables on SMALL and 80 query tables on LARGE, 
for which we report the average, 10th, and 90th percentile query times.
Notice that the query time can vary depending on the complexity of the query table.
\SANTOS average query times are faster than $D^3L$ query time, since $D^3L$ searches through each of their five indexes when finding unionable tables, whereas we only search through synthesized KB index, YAGO index, or both. Also, the query time of $D^{3}L$ and \SANTOS are comparable on SMALL because $D^{3}L$ is faster on smaller data lakes. However, \SANTOS is faster than $D^{3}L$ for larger data lakes as suggested by the average query time on TUS (over 3 times faster) and LARGE (almost 6 times faster).
All in all, even with the new synthesized KB, our indexing and query times are comparable or even faster than those of a state-of-the-art approach.

\section{Conclusion}
We presented \SANTOS, a method for finding unionable tables in data lakes based on both column and relationship semantics.  
\SANTOS discovers and uses
relationship semantics between pairs of columns in a table using an existing knowledge base (KB) and a synthesized KB created by exploiting the knowledge of the data lake.
We conducted experiments on an adapted version of the existing TUS benchmark as well as our new SMALL and LARGE benchmarks and showed that \SANTOS unionability search outperforms a state-of-the-art table union approach.
Also, the experimental results showed the robustness of our approach and the importance of our synthesized KB in overcoming curated KBs with limited coverage. 
\smallbreak
\noindent
\introparagraph{Acknowledgements}
This work was supported in part by NSF
under award numbers IIS-1956096, IIS-2107248 and IIS-1762268.
\bibliographystyle{ACM-Reference-Format}
\bibliography{sample}


\end{document}